\documentclass[prd, preprint,nofootinbib,superscriptaddress,eqsecnum,12pt]{revtex4-1}
\pdfoutput=1

\usepackage{paralist}
\usepackage{amsmath}
\usepackage{xcolor}
\usepackage{geometry}
\usepackage{amssymb,mathrsfs}
\usepackage{hyperref}
\usepackage{gensymb}
\usepackage{slashed}
\usepackage{graphicx,epsfig,dcolumn,bm,xspace}
\usepackage{url}
\usepackage{tikz}
\usepackage{cancel}
\usepackage[compat=1.1.0]{tikz-feynman}

\hypersetup{colorlinks,citecolor= red ,linkcolor= niceblue, urlcolor=niceblue}
\definecolor{niceblue}{rgb}{0.1,0.2,0.6}

\newcounter{mycountit}
\newcommand{\begmyen}{\setcounter{mycountit}{1}}
\newcommand{\myitem}{$\bullet$~}
\addtocounter{mycountit}{1}

\newcommand{\D}{\mathrm{d}}

\newcommand{\gmu}{\gamma^{\mu}}

\newcommand{\gMu}{\gamma_{\mu}}

\newcommand{\gfive}{\gamma^5}

\newcount\rem \rem=0
\ifnum \rem=1
\newcommand{\ygn}[1]{}
\newcommand{\wt}[1]{}
\newcommand{\mign}[1]{}
\newcommand{\margn}[1]{}
\else
\newcommand{\ygn}[1]{{\bf \color{red} [YG:~#1]}}
\newcommand{\wt}[1]{{\bf \color{blue} [WT:~#1]}}
\newcommand{\mign}[1]{{\bf \color{green} [Mijo:~#1]}}
\newcommand{\margn}[1]{{\bf \color{orange} [MR:~#1]}}
\fi
\newcommand{\nnu}{n_{\nu}}
\newcommand{\na}{n_A}
\newcommand{\nga}{n_{\Gamma}}
\newcommand{\nphi}{n_{\phi}}
\newcommand{\npsi}{n_{\psi}}

\def\beq{\begin{equation}}
\def\eeq{\end{equation}}

\begin{document}

\title{Fermion pair radiation from accelerating classical systems}
\author{Margarita Gavrilova}
\email{mg2333@cornell.edu}
\affiliation{Department of Physics, LEPP, Cornell University, Ithaca, NY 14853, USA}
\author{Mitrajyoti Ghosh}
\email{mg2338@cornell.edu}
\affiliation{Department of Physics, LEPP, Cornell University, Ithaca, NY 14853, USA}
\author{Yuval~Grossman}
\email{yg73@cornell.edu}
\affiliation{Department of Physics, LEPP, Cornell University, Ithaca, NY 14853, USA}
\author{Walter Tangarife}
\email{wtangarife@luc.edu}
\affiliation{Department of Physics, Loyola University Chicago, Chicago, IL 60660, USA}
\author{Tien-Hsueh Tsai}
\email{s106022901@m106.nthu.edu.tw}
\affiliation{Department of Physics, National Tsing Hua University, Hsinchu 300, Taiwan}

\begin{abstract}
Accelerating classical systems that couple to a fermion-antifermion pair at the microscopic level can radiate pairs of fermions and lose energy in the process. In this work, we 
derive the generalization of the Larmor formula for fermion pair radiation. 
We focus on the case of a point-like classical source in an elliptical orbit that emits fermions
through vector and scalar mediators. 
Ultra-light fermion emission from such systems becomes relevant when the mass of the mediator is larger than the frequency of the periodic motion.
This enables us to probe regions of the parameter space that are inaccessible in on-shell bosonic radiation.
We apply our results to pulsar binaries with mediators that couple to muons and neutrinos. Using current data on binary period decays, we extract bounds on the parameters of such models.

\end{abstract}

\maketitle
\tableofcontents
\section{Introduction} \label{sec:intro}

Radiation by a classical system is a well-known phenomenon.
Probably the most familiar example is the radiation of electromagnetic waves by an accelerating point-like particle. The power loss, in this case, is calculated using the famous Larmor formula~\cite{LarmorLXIIIOT,Jackson:1998nia}, which, in natural units, is given by
\beq \label{eq:larmor}
P_{\text{loss}} = \frac{1}{6 \pi} q^2\bm{a}^2,  
\eeq
where $q$ is the electric charge of the particle and $\bm{a}$ is its acceleration. The Larmor formula in Eq.~\eqref{eq:larmor} has been also generalized to other types of radiation by accelerating classical sources, such as radiation of massive vector and scalar bosons~\cite{Krause:1994ar,Mohanty:1994yi,Dror:2019uea, KumarPoddar:2019ceq,Huang:2018pbu,KumarPoddar:2019jxe,Hook:2017psm,Mohanty:1994yi}. 

Generalizations of the Larmor formula to exotic types of radiation are motivated, among other things, by their applications to new physics searches. The basic idea is that if a new physics radiation accompanies an accelerating astrophysical object, the power loss effect can be enhanced thanks to the large number density of an object, even if the coupling between the new physics and the Standard Model (SM) is very small. This expected enhancement can be used to obtain constraints on various new physics scenarios using astrophysical observations. One example is the radiation of an ultra-light gauged $L_{\mu}-L_{\tau}$ vector boson~\cite{Foot:1990mn,He:1991qd,Foot:1994vd, Heeck:2011wj}
by pulsar binaries.
The measurement of the orbital period decay, when compared to the prediction due to the gravitational wave (GW) radiation, was used to constrain the mass of the $L_\mu - L_\tau$ gauge boson and its couplings to the SM~\cite{KumarPoddar:2019ceq,Dror:2019uea, Kopp:2018jom,Alexander:2018qzg,Choi:2018axi, Fabbrichesi:2019ema, Seymour:2019tir}.

In this paper, we extend the previous work and derive the generalization of the Larmor formula to the case of fermion-antifermion pair radiation by classical systems.
The interest in this scenario is twofold. First, it is interesting theoretically since it is one more example of a case where a fermion pair behaves like a boson (other cases  are Cooper pairs in superconductors and the mediation of forces between objects via 2-fermion forces \cite{Feinberg:1968zz,Hsu:1992tg,Ghosh:2019dmi,Ghosh:2022nzo}). Thus we can study the coherent radiation of fermions. The key point is that single-fermion emission changes the source and thus can not be treated classically. Fermion-pair emission, however, can take place without changing any quantum degrees of freedom of the emitting system (such as spin). Thus, fermion-pair emission (or emission of any even number of fermions) can be treated classically.

The second aspect is phenomenological. In particular, we consider radiation by astrophysical systems. In the SM, as we show below, the effect of the fermion pair radiation is negligible. In beyond the SM (BSM) theories, however, such processes can be enhanced, enabling us to probe various new physics scenarios using astrophysical observations.
In particular, fermion-pair radiation can become significant in models with a new light mediator (a vector or scalar boson) that couples to some light fermionic degrees of freedom. These fermionic degrees of freedom can be the well-known neutrinos or some new BSM fermions. The effects of this  radiation can become relevant when the mediator is too heavy to be produced on-shell, but the fermions are much lighter and can be radiated out. Since fermion pairs can be produced via off-shell mediators, the fermion pair radiation can be used to probe broader regions of the parameter space of such models.

As a particular application of our result for the fermion-pair radiation, we consider two models: (\emph{i}) a model with a gauged  $L_\mu-L_\tau$ symmetry and (\emph{ii}) a model with a muonophilic scalar that couples to the muon and the muon neutrino. We study the implications of these scenarios for the power loss by pulsar binaries and compare our results to the cases of on-shell vector boson radiation~\cite{KumarPoddar:2019ceq,Dror:2019uea,Krause:1994ar} and on-shell scalar radiation~\cite{Krause:1994ar}. A stark difference is that the emission of neutrino pairs in a particular harmonic mode of the periodic system is not kinematically forbidden when the mediator mass becomes larger than the frequency of that particular mode. In the case of on-shell bosonic radiation, radiation from a harmonic mode is cut off once the boson mass exceeds the frequency of that particular mode due to energy conservation. We use the available period decay data for pulsar binaries 
to demonstrate how neutrino pair radiation, mediated by BSM bosons, can be used to probe a broader parameter space than the on-shell boson emission. We, however, do not perform a comprehensive study of other bounds on the models we consider.

This paper is organized as follows: 
In Sec.~\ref{sec:formalism}, we discuss the general machinery required for calculating fermion-pair radiation from a classical system. In Sec.~\ref{sec:discformula}, we discuss the main features of the power-loss formula. In Sec.~\ref{sec:Rad_NS}, we perform the computation for the particular case where the classical system is a binary system. We then use available data to place constraints on the parameters of a few models. We conclude in Sec.~\ref{sec:conc}. The detailed calculations are shown in the appendix.

\section{Fermion pair radiation by a point-like object 
}\label{sec:formalism}
In this section, we outline the calculation of the power of fermion-pair radiation that accompanies a non-relativistic point-like object. We formulate a general approach to the derivation of the power loss formula with a focus on the case of elliptical orbits. The fermion pair radiation is realized in our analysis via the coupling of the classical object to a massive boson: a vector, or a scalar, which is unstable and decays into a fermion pair. We consider the emission of Dirac fermions and generalize our result to the case of Weyl fermions when we discuss the application of our result to the SM in Section~\ref{sec:fermion-pair-rad-SM}. While a point-like object is a purely theoretical entity, it is worthwhile to perform this calculation since the approximation of a radiating extended object as a point is valid in the limit of long-wavelength radiation.

\subsection{General formalism}\label{sec:gen-form}

We describe a point-like object as a classical source using classical current, $J_{\text{cl}} ^{\mu} (x)$  and classical density, $\rho_{\text{cl}} (x)$, which are given by
\beq
J_{\text{cl}} ^{\mu} (x) = Q \delta^3 ({\bm x} - {\bm x}(t) ) u ^ {\mu}, \label{eq:jcldef-ellipse}\eeq
\beq \rho_{\text{cl}} (x) = N \delta^3 ({\bm x} - {\bm x}(t)). 
\label{eq:rhocldef-ellipse}
\eeq
Here, $Q$ is the total charge of the object under the symmetry of interest, $N$ is the number of the relevant microscopic constituents, ${\bm x} (t)$ is its position as a function of time, $t$, and $u^{\mu}$ is its four-velocity. 

Assuming motion in the $x-y$ plane, in the non-relativistic limit, the  four-velocity of the object is given by 
\begin{equation}\label{eq:umu}
    u^{\mu} = \left( 1, \dot x, \dot y, 0\right).
\end{equation}
We focus on the case of the elliptical motion in the $x-y$ plane, which can be parametrically described by
\beq\label{eq:ellipse-param}
x = a(\cos \xi -e) , \qquad y = a \sqrt{1-e^2} \sin \xi , \qquad \Omega \, t = \xi - e \sin \xi,
\eeq
where $e$ is the eccentricity, $a$ is the semi-major axis of the ellipse, and $\Omega$ is the fundamental frequency of revolution. One full revolution around the ellipse corresponds to changing the parameter $\xi$ from $0$ to $2\pi$.

The power loss due to the fermion-pair radiation is calculated using
\beq P_{\text{loss}} = \int  (\omega_{1} +\omega_{2} )\, \D \Gamma, \label{eq:plossgen}
\eeq
where $\omega_1$ and $\omega_2$ are the energies of the emitted fermion and anti-fermion, respectively, and $\D \Gamma $ is the differential rate of the fermion-pair emission. The rate depends on the type of mediator, i.e., a scalar or a vector, and the specific form of the classical current or density. 

\begin{figure}[t]
\centering
\begin{tikzpicture}
\begin{feynman}
\node [crossed dot] (b);
\vertex [ right=of b] (c);
\vertex [above right=of c] (f2) {\(\psi, k_1\)};
\vertex [below right=of c] (f3) {\(\overline \psi, k_2\)};
\diagram* {
(b)-- [scalar, edge label'=~~mediator~~] (c),
(c) -- [anti fermion] (f2),
(c) -- [fermion] (f3),
		};
\end{feynman}
\end{tikzpicture}
\caption{Feynman diagram for a fermion pair emission by a classical current.}
    \label{fig:ffemission}
\end{figure}
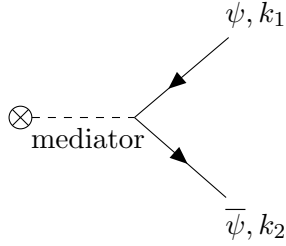

In general, the acceleration is not constant. In the case of periodic orbits, the motion can be decomposed into harmonic modes with frequencies $\Omega_n = n \Omega$, where $\Omega$ is the fundamental frequency of revolution. The total emission rate can then be written as a sum of emission rates at different harmonics $n$,
\begin{equation} 
    \D \Gamma = \sum_{n} \D \Gamma_n\,.
\end{equation}
The sum goes over all kinematically allowed harmonics $n > 2m_\psi/\Omega$, where $m_\psi$ is the mass of the emitted fermions. The emission rate at harmonic $n$ is found using
\beq \label{eq:dGamma2}
\D \Gamma_n = \sum_{s_1,s_2}|{\cal M}_n (s_1 ,s_2)|^2(2\pi)\delta(\Omega_n-\omega_1-\omega_2)\frac{\D ^3{{\bm k}_1}}{(2\pi)^3 \omega_1}\frac{\D ^3{{\bm k}_2}}{(2\pi)^3 \omega_2}.
\eeq
Here, $k_1 = \left(\omega_1, \bm{k}_1\right)$ and $k_2 = \left(\omega_2, \bm{k}_2\right)$ are the four-momenta of the fermion and anti-fermion respectively, and $s_1$($s_2$) is the spin of the fermion (anti-fermion). The microscopic physics enters via ${\cal M}_n\left(s_1,s_2\right)$, which is the matrix element of the fermion-pair emission at harmonic $n$. At leading order, this matrix element is obtained from the diagram in Fig.~\ref{fig:ffemission}. In the diagram, $\otimes$ denotes the classical source, which is given by the classical current, $J_\text{cl}^\mu(x)$, in the case of  vector mediator and by the density, $\rho_\text{cl}(x)$, in the case of the scalar mediated radiation.
The total power loss via fermion-pair radiation is simply a sum of power losses over all harmonics
\beq  P_{\rm loss} = \sum_n  P_n, \qquad P_n =  \int (\omega_1 + \omega_2)\, \D \Gamma_n. \label{eq:plosssummary} \eeq
Here, $P_n$ is the power loss of the $n$th harmonic.

In what follows, we consider two types of mediators: a massive gauge boson and  a massive scalar. We only consider $s$-channel exchange and remark on $t$-channel exchange at the end of this subsection.

First, we consider a vector mediator, $A_\mu$, that corresponds to a broken $\operatorname{U}(1)^\prime$ and has mass $m_A$. This gauge boson couples to a classical current $J^\mu_\text{cl}(x)$, which has charge $Q$ under $\operatorname{U}(1)^\prime$. The gauge boson $A_\mu$ is unstable and decays into a fermion pair. The terms in the effective Lagrangian, relevant for the fermion-pair radiation via $A_\mu$, are
\begin{equation}\label{eq:L-vec-int} 
{\cal L}_\text{eff} \supset g A_{\mu} J_{\text{cl}}^{\mu} +  g q_{\psi} \bar{\psi} \gmu A_{\mu} \psi \, ,
\end{equation}
where $q_{\psi}$ is the $\operatorname{U}(1)^\prime$ charge of the fermion $\psi$, $g$ is a dimensionless coupling constant, and $J_{\text{cl}}^{\mu}(x)$ is the classical current defined in Eq.~\eqref{eq:jcldef-ellipse}. Both the vector boson and the fermions are assumed to be massive with masses $m_A$ and $m_\psi$, respectively. The leading order matrix element for the emission, at the $n-$th harmonic, is given by
\beq \label{eq:M-vector}
{\cal M}_n(s_1,s_2) =g^2 q_{\psi}\,\bar{u}(k_1,s_1)\gamma^\mu v(k_2,s_2)\,\frac{i(-\eta_{\mu \nu}+(k_1+k_2)_{\mu}(k_1+k_2)_{\nu}/m_A^2)}{(k_1+k_2)^2-m_A^2 + i m_A \Gamma_A }\, J_{\text{cl}}^{\nu}(\Omega_n)\, ,
\eeq
where $J^{\nu}_{\text{cl}} (\Omega_n)$ is the Fourier transform of $J_{\text{cl}}^{\nu} (x)$, given by
\begin{equation}
    J^{\nu}_{\text{cl}} (\Omega_n) = {\Omega \over 2 \pi} \int_0^{2\pi/\Omega}  \D t \int \D ^3 \bm{x} \  e^{i(  n \Omega t - \bm{p}\cdot \bm x)} J^{\nu}_{\text{cl}} (x)
\end{equation}
with $\bm{p} = \bm{k}_1 + \bm{k}_2$, $\Gamma_A$ is the decay width of the gauge boson, and $2\pi/\Omega$ is the period. We assume that the decay into a $\bar \psi \psi$ pair is the only decay channel for the gauge boson $A_\mu$, and that the fermion mass $m_\psi$ is negligible compared to the gauge boson mass $m_A$. Under these assumptions, the decay width of $A_\mu$ is given by
\beq
\Gamma_A = {g^2q_\psi^2 m_A \over 12 \pi}.
\eeq

The other case we consider is that of a scalar mediator, $\phi$, for which the relevant terms in the Lagrangian are
\beq {\cal L} \supset g \phi \rho_{\text{cl}} +  g^\prime \phi \bar{\psi}  \psi,
\eeq
where $g$ is the dimensionless coupling between the scalar $\phi$ and the classical source, $g^\prime$ is the Yukawa coupling of the fermion $\psi$ to the scalar $\phi$, and $\rho_{\text{cl}}(x)$ is the number density of relevant particles in the classical source. Both the scalar and the fermions are assumed to be massive with masses $m_\phi$ and $m_\psi$, respectively. The matrix element in this case is given by
\beq \label{eq:M-scalar}
	{\cal M}_n(s_1,s_2) =g g' \bar{u}(k_1,s_1) v(k_2,s_2)\frac{i \rho_{\text{cl}} (\Omega_n)}{(k_1+k_2)^2-m_{\phi}^2 + i m_{\phi} \Gamma_{\phi}} ,
\eeq
where $\rho_{\text{cl}} (\Omega_n)$ is the Fourier transform of $\rho_{\text{cl}}(x)$,
\beq \rho_{\text{cl}} (\Omega_n) = {\Omega \over 2 \pi} \int_0^{2\pi/\Omega}  \D t \int \D ^3 \bm{x} \  e^{i(  n \Omega t - \bm{p}\cdot \bm x)} \rho_{\text{cl}} (x),\eeq
and the decay width of the scalar is $\Gamma_{\phi}$.
As in the case of the vector mediator, we assume that the fermionic decay mode is the only available mode, and the fermion mass $m_\psi$ can be neglected compared to the mass of a scalar $m_\phi$. Thus we have 
\begin{equation}
    \Gamma_{\phi} = \frac{g^{\prime 2} m_{\phi}}{8 \pi}.
\end{equation}

So far, we have only considered the $s$-channel contribution to the fermion pair radiation. Fermion pair radiation via $t-$channel process mediated by a vector or scalar is also a possibility. Such contributions, however, are highly suppressed for $m_S \gg \Omega, m_M$, where $m_S$ is the mass of the particles in the source that couple to the fermion pairs $\bar \psi \psi$ at the microscopic level, and $m_M$ is the mediator mass.
Since the emitted fermions have energy of the order of $\Omega$, the fundamental frequency of the system, the $t$-channel contribution to the momentum entering the propagator is of the order of $m_S - \Omega$. Thus the $t$-channel propagator is schematically given by
\beq 
\Pi \sim {1 \over (m_S - \Omega)^2 - m_M ^2}.
\eeq
In the case  where $m_S$ is much larger than both $\Omega$ and $m_M$, the propagator is dominated by the mass of the source particles, and the process is heavily suppressed. In this paper, we assume that the mass hierarchy $m_S \gg \Omega, m_M$ and neglect the $t-$channel contributions to the fermion pair radiation everywhere. 

\subsection{Power loss formulae} \label{sec:plossformulae}
Using Eqs.~\eqref{eq:dGamma2}--\eqref{eq:M-scalar}, we can calculate the power loss via fermion-pair radiation from a point-like object moving in an elliptical orbit. 
The detailed derivations are shown in Appendix~\ref{app:calc}, and here we only quote the final result. The power loss due to fermion-pair emission in harmonic $n > 2 m_{\psi}/\Omega$, for the cases of the vector and scalar mediator, can be written as
\begin{align} 
P^{A}_n &= \frac{g^4 q_{\psi}^2Q^2 }{12 \pi^3} \,a^2 \Omega^4\,  
 B_n^A(n_A,n_\psi,n_\Gamma), \label{eq:Ploss-A-mai} \\
P^{\phi}_n &= \frac{g^2 g'^2 N^2}{12 \pi^3}\, a^2 \Omega^4\,
B_n^\phi(n_\phi,n_\psi,n_\Gamma).\label{eq:Ploss-phi-mai}
\end{align}
The functions $B_n^M (n_A, n_\psi, n_\Gamma)$, where $M = A, \phi$, are given by
\beq \label{eq:Bn-mai}
B_n^M(n_M,n_\psi,n_\Gamma)\,\equiv\,\left(J^\prime_n(ne)^2 + \frac{1 - e^2}{e^2}J_n(ne)^2\right) \int_{n_\psi}^{n- n_\psi} \D x \  F^{M}(x,n,n_M,n_\psi,n_\Gamma).
\eeq
Here 
\beq
n_M \equiv m_M/\Omega, \qquad n_\psi \equiv m_{\psi}/\Omega, \qquad
n_\Gamma \equiv \Gamma_M/\Omega,
\eeq
and $J_n(ne)$ is a Bessel function of order $n$ with argument $ne$.
The integration variable in Eq.~\eqref{eq:Bn-mai} is defined by $x \equiv \omega_1/ \Omega $, where $\omega_1$ is the energy of one of the final-state fermion. In what follows, for brevity, we use the notation 
\beq
F^M(x) \equiv F^M(x,n,n_M, n_\psi, n_\Gamma), \qquad
B_n^M \equiv B_n^M(n_M,n_\psi, n_\Gamma).
\eeq
The functions $F^M(x)$ have the general form
\begin{eqnarray}
\label{eq:FM-gen-mai}
F^M(x) = F^M_0 (x) &+&  {F^M_1(x)\over n_M \nga} \left[ \tan^{-1} \left({a(x)+b(x) \over n_M \nga }\right) - \tan^{-1} \left( {a(x)-b(x) \over n_M \nga }\right)\right] \nonumber \\
&+& F^M_2(x) \tanh ^{-1} \left ({2 a(x) b(x) \over  a(x)^2 + b(x)^2 + n_M^2 \nga^2}\right), \label{eq:fx}
\end{eqnarray}
with $a(x)$ and $b(x)$ being universal for both gauge boson and scalar mediators,
\begin{eqnarray}\label{eq:a-b-defs-mai}
a(x) &=& 2 n_\psi ^2 -n_M^2 + 2 n x - 2 x^2\,, \nonumber\\
b(x) &=& 2\sqrt{x^2 - n_\psi^2}\sqrt{(n-x)^2-n_\psi^2}\, .
\end{eqnarray}
The functions $F_0^M(x)$, $F_1^M(x)$, and $F^M_2(x)$ are different for the two cases. For a gauge boson mediator, we obtain
\begin{eqnarray}
F^A_0(x) &=& b(x)/2n \, ,\nonumber \\
F^A_1(x)  &=& {1 \over 4n} \left(\na^4 + 4n^2 n_\psi^2 - \na^2 \nga^2 + 2 \na^2 n^2 - 4nx\na^2 + 4x^2 \na^2 \right) \nonumber\, , \\
F^A_2(x) &=& {1 \over 2n} \left( \na^2 +n^2 -2 n x + 2x^2\right)\, , \label{eq:f0f1f2b-mai}
\end{eqnarray}
while for a scalar mediator,
\begin{eqnarray}\label{eq:f0f1f2scal-mai}
F^\phi_0 (x) &=& -b(x)/2n \, ,\nonumber \\
F^\phi_1(x)  &=& {1 \over 4n} \left(\nphi^2 \nga^2 +(n^2-\nphi^2)(\nphi^2-4n_\psi^2) \right) \, ,\nonumber \\
F^\phi_2(x) &=& {1 \over 4n} \left( n^2+4 n_\psi^2-2\nphi^2 \right) .
\end{eqnarray}

Eqs.~\eqref{eq:Ploss-A-mai}--\eqref{eq:f0f1f2scal-mai} are the main results of our work. Analytical integration of $F^A(x)$ and $F^\phi(x)$ is challenging, but it still can be performed in certain limits. In Sec.~\ref{sec:asympt-beh}, we consider two limiting cases: the case of $n_M \ll 1$, which reproduces the Larmor formula, and $n_M \gg 1$, which is relevant for the fermion pair radiation in the SM. In general, however, calculating the power loss requires numerical analysis.  We perform such an analysis in Sec.~\ref{sec:Rad_NS} when we discuss a particular phenomenological application of our result.

\section{Discussion of the power-loss formula} \label{sec:discformula}

The power loss due to fermion-pair emission by a classical source on an elliptical orbit is given by Eqs.~\eqref{eq:Ploss-A-mai}-\eqref{eq:f0f1f2scal-mai}. Below we discuss the main features and the asymptotic behavior of this result.

\subsection{General features of the power-loss formula}
We start with the general features that hold for  both the vector and scalar cases.

\begmyen

\myitem
The radiation rate is proportional to the  charge-squared; that is, the functions $P^A_n$ and $P^\phi_n$ 
depend on $Q^2$ and $N^2$, respectively. This is a manifestation of the fact that the fermion-pair radiation that we are considering is coherent.

\myitem 
The form of $F^M (x)$, with $M = A, \phi$, in Eq.~\eqref{eq:FM-gen-mai} is somewhat general. We show in Appendix~\ref{app:calc} that the overall form of $F^M(x)$, at the tree level, is the same for any renormalizable theory that couples fermions to a classical source moving in an elliptical orbit. Note that the functions $a(x)$ and $b(x)$ defined in Eq.~\eqref{eq:a-b-defs-mai} are purely kinematic and thus have the same form for any theory of fermion pair emission, while the form of $F_0^M(x), F_1^M(x)$, and $F_2^M(x)$ vary with the theory considered. For instance, considering non-renormalizable interactions would lead to a different momentum dependence of the matrix element that could, in principle, change the form of $F^M(x)$.

\myitem
The power loss for both vector and scalar mediators behaves qualitatively the same way despite the different functional forms of $F_i^A(x)$ vs. $F_i^\phi(x)$, with $i=0,1,2$. This is not surprising since there is nothing fundamentally different between the matrix elements for the vector and scalar cases. 

\myitem
Energy conservation implies that the functions $F^M(x)$ are invariant under   $x \to (n-x)$ exchange. The reason is that the total energy radiated in fermion pairs in the $n$-th harmonic is $n\Omega$. The transformation $x \to (n-x)$ exchanges the energies of the emitted fermion and anti-fermion, and the emission rate is the same regardless of the order in which the integrals are carried out. This invariance results from the fact that the fermion-antifermion emission from a classical system is essentially a 2-body decay. Note that this has nothing to do with the details of the considered model.

\myitem
For $\na <n$, the power loss has a very weak dependence on $\na$. This is true for the particular models that we chose here but is not expected to be true in general. For an example when this is not the case, see the discussion of Proca fields in Ref.~\cite{KumarPoddar:2019ceq}, where dependence on $\na$ appears due to the absence of gauge symmetry.

\myitem
There is an interplay of three energy scales: The mass of the mediator, $m_M$, the mass of the fermion, $m_\psi$, and the frequency of the harmonics, $n \Omega$. The fermions cannot be produced when $2m_{\psi}>n\Omega$. In the opposite limit, when $2m_{\psi}<n\Omega$, the production rate depends strongly on the mediator mass.
For $m_M<2m_{\psi}<n\Omega$, fermion production is strongly suppressed since the on-shell boson is kinematically forbidden from decaying into fermions. (Note that strictly speaking, our result cannot be straightforwardly applied in this case as everywhere we assume $\Gamma_M > 0$.) For $2m_{\psi}<m_M<n\Omega$, the fermions are produced via decay of the on-shell mediator. Thus the power loss in the fermion-pair radiation is equal to that of the on-shell boson radiation. The region of the parameter space where $m_M>n\Omega> 2m_{\psi}$ is of the most interest to us, as in this region the fermions are kinematically allowed, the mediator is off-shell, and therefore the fermion pair emission is most significant.

\begin{figure}[t]
\includegraphics[scale=0.65]{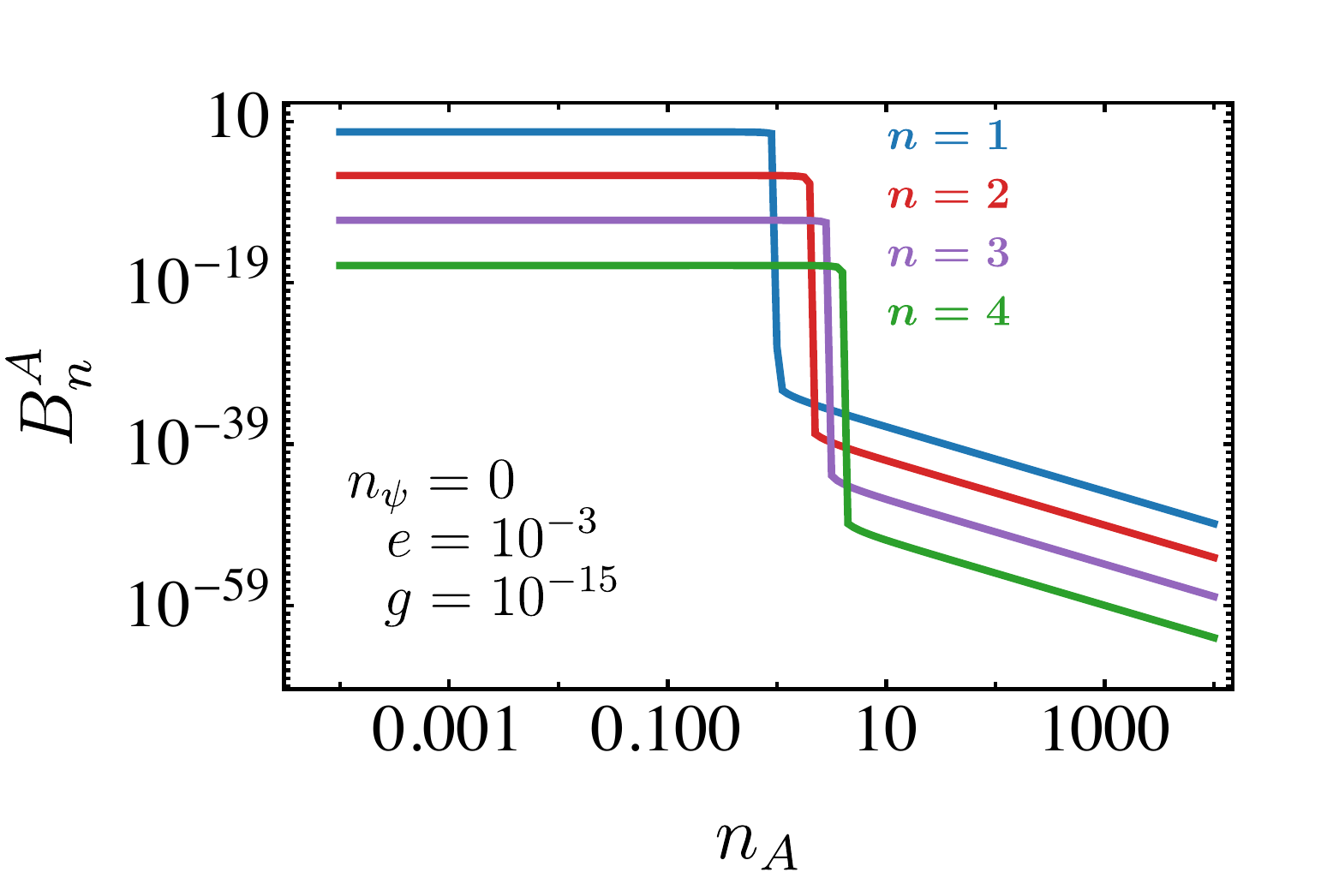}
\caption{$B^A_n~{\rm vs}~n_A$ for fixed eccentricity, $e = 10^{-3}$, coupling constant $g = 10^{-15}$, and massless final state fermions, $m_\psi=0$. See Eqs.~\eqref{eq:Bn-mai}-\eqref{eq:f0f1f2b-mai} for the definition of $B_n^A$.}
\label{fig:pvsna}
\end{figure}

\myitem
As an example that illustrates the qualitative features of the power loss, consider Fig.~\ref{fig:pvsna}. It shows $B^A_n$, defined in Eq.~\eqref{eq:Bn-mai}, as a function of $n_A$ for massless fermions for the first four harmonics. The most striking feature of the plots is a sharp drop at $n_A \sim n$. This behavior follows from the fact that at $n_A \sim n$, the radiation regime switches from the radiation dominated by on-shell boson production ($n_A < n$), which is proportional to $g^2$ to the off-shell production ($n_A > n$) proportional to $g^4$. The power loss in the regime dominated by fermion-pair radiation is thus suppressed by $g^2$ compared to the power loss in the regime dominated by the on-shell boson radiation. The power loss in the case of the scalar mediator exhibits the same behavior.

\myitem Comparing our results to the cases of vector~\cite{KumarPoddar:2019ceq, Dror:2019uea, Krause:1994ar} and scalar radiation~\cite{Krause:1994ar}, we note that from kinematic considerations alone, boson radiation drops to zero as soon as $n_M = n$. This is not what we observe for the fermion-pair emission. In the case of fermion-pair radiation, off-shell boson production is possible, even though there is an extra suppression by $g^2$ for a vector and ${g^\prime}^2$ for a scalar compared to on-shell boson radiation. As a result, the regime $n_M > n$ opens up new regions of the parameter space for each harmonic $n$ and is of particular phenomenological interest to us.

\myitem
Next, we remark on the dependence of the power loss on the eccentricity in the case of orbits close to circular. For that, we note that the eccentricity only enters the power loss through the Bessel function prefactor of $B^M_n$ in Eq.~\eqref{eq:Bn-mai}, which we denote as $K(n,e)$,
\begin{equation}
    K(n,e) = J^\prime_n(ne)^2 + \frac{1 - e^2}{e^2}J_n(ne)^2\,.
\end{equation}
We recall that $J_n(z)$ and $J_n^\prime(z)$ behave asymptotically, in the limit $z\ll 1$, as
\beq\label{eq:Bess-assympt}
J_n(z) \approx \frac{1}{n!}\left(\frac{z}{2}\right)^n, \qquad J_n^\prime(z) \approx \frac{n}{n!}\frac{1}{2}\left(\frac{z}{2}\right)^{n-1} \approx \frac{n}{z} J_n(z), \qquad z \ll 1.
\eeq
Using Eq.~\eqref{eq:Bess-assympt}, we find for the eccentricity dependent prefactor $K(n,e)$, in the limit $ne\ll 1$, that
\begin{align}\label{eq:Kne-assympt}
K(n,e)& = J^\prime_n(ne)^2 + \frac{n^2 - (ne)^2}{(ne)^2}J_n(z)^2 \approx J^\prime_n(ne)^2 + \frac{n^2}{(ne)^2} J_n(z)^2 \nonumber \\ & = 2\frac{n^2}{z^2}J_n(ne)^2 = \frac{(ne)^{2n-2}}{2^{2n-1}}\frac{n^2}{\left(n!\right)^2} = \frac{(ne)^{2n-2}}{2^{2n-1}\left((n-1)!\right)^2}.
\end{align}
Thus we learn that in the limit $ne \ll 1$, prefactor $K(n,e)$ scales with the eccentricity as
\begin{equation}\label{eq:Pn-e}
    K(n,e) \propto \left(ne\right)^{2n-2}.
\end{equation}
This shows that for small eccentricities (and thus orbits close to circular ones), the contributions from higher harmonics die away very fast as $n$ increases. For $n=1$ and $e\ll 1$, we have $K(1,e)\approx 1/2$. For each subsequent harmonic power drops by a factor of order $e^2$, until the factorial in the denominator of $K(n,e)$ (see Eq.~\eqref{eq:Kne-assympt}) starts to dominate. Then the contributions from the higher harmonics start to decay away even faster. Fig.~\ref{fig:pvsna} illustrates the behavior of the power loss for the first four harmonics in the case of small eccentricity $e=10^{-3}$.

\begin{figure}[t]
\includegraphics[scale=0.52]{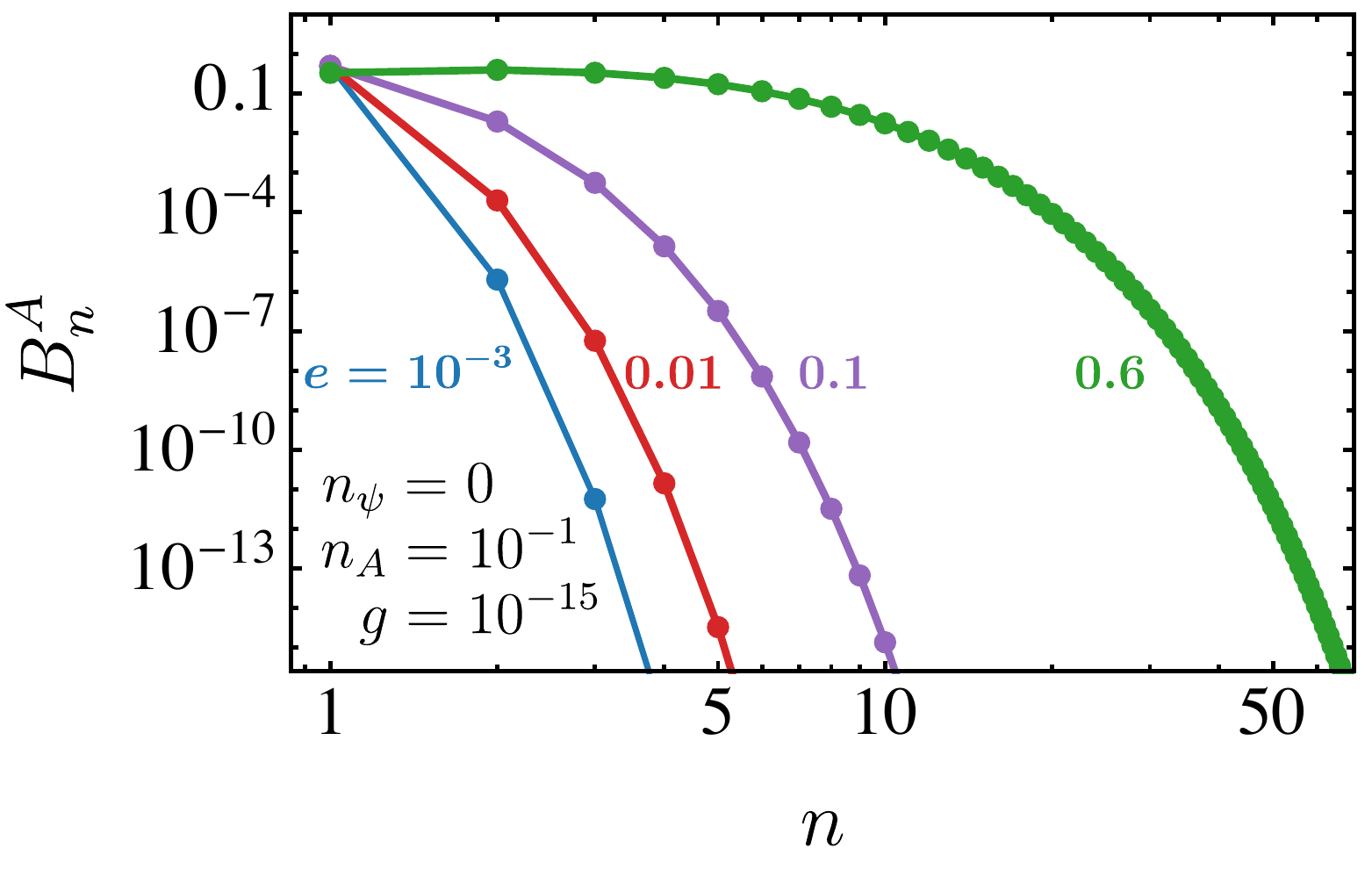} \hspace{5pt}
\includegraphics[scale=0.52]{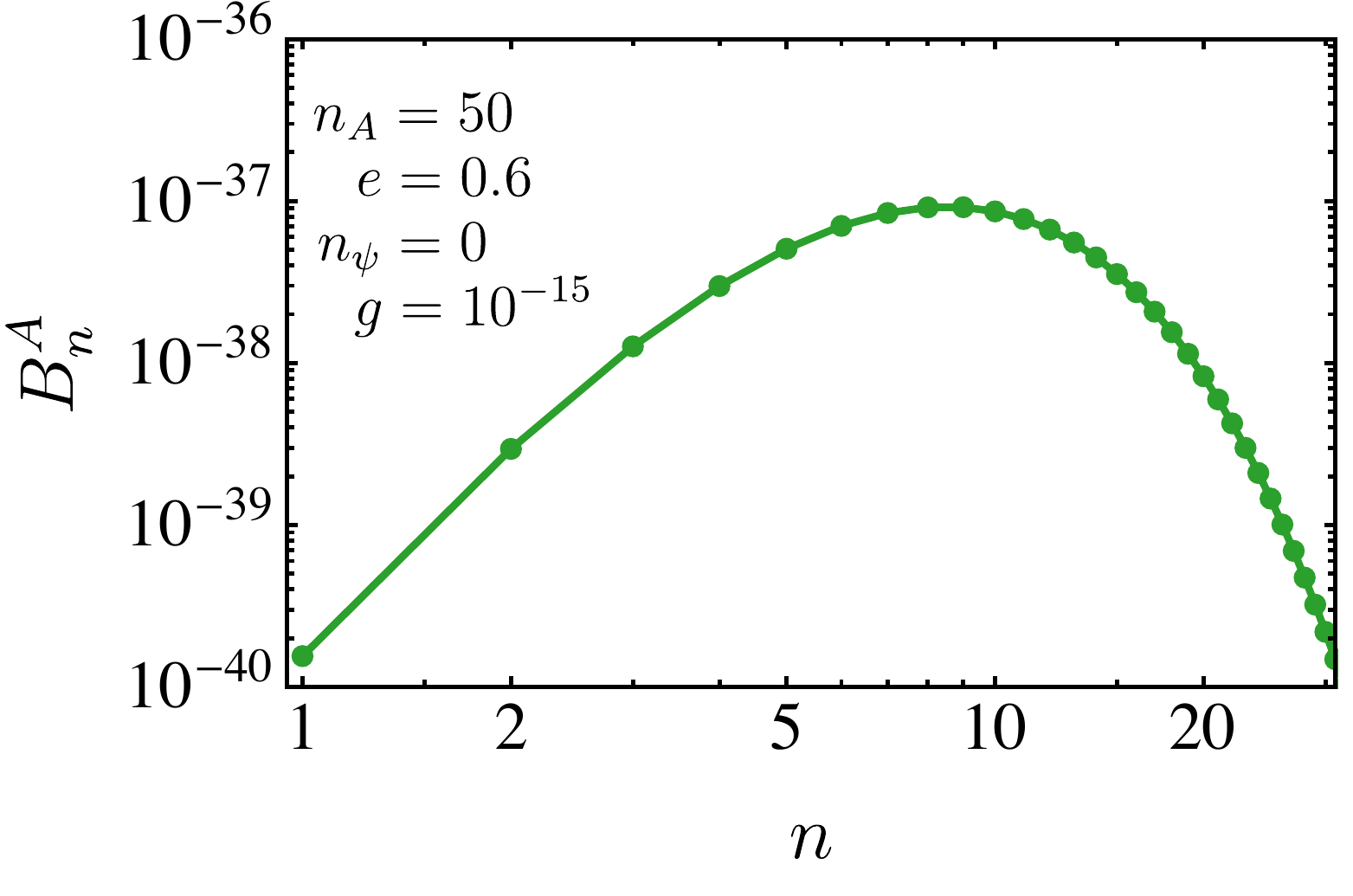}
\caption{\emph{Left:} $B_n^A$ as a function of $n$ in the regime where the radiation is dominated by on-shell boson production. Different colors correspond to different values of eccentricity. The values of $n_\psi$, $n_A$ and $g$ are fixed. \emph{Right:}  $B_n^A$ as a function of $n$ for a highly eccentric orbit with $e=0.6$ in the regime where the radiation is dominated by off-shell boson production.
}
\label{fig:BnA-vs-n}
\end{figure}

\myitem 
The case of highly eccentric orbits $e\sim 1$ is significantly more involved. First, the contributions from different modes do not follow the simple hierarchy of the low eccentricity case. The contributions from higher modes can be of the same order or even larger than the first mode depending on the values of other parameters. See the left panel of Fig.~\ref{fig:BnA-vs-n} to compare the $n$-dependence of $B_n^A$ for different eccentricity values. Second, as Fig.~\ref{fig:BnA-vs-n} demonstrates, the hierarchy of modes in the on-shell dominated part of the parameter space does not carry into the off-shell dominated region. Consider the green line corresponding to a highly eccentric orbit with $e=0.6$. For $n_A=10^{-1}$ (left panel), the maximum contribution to the power loss comes from the mode with $n=2$ and the first 5 modes contribute at about the same order. The situation is drastically different for $n_A = 50$ (right panel). The maximum contribution to the power loss comes from the $n=8$ mode. We learn that for $e \sim 1$, generally speaking, the power loss per mode first increases as we increase $n$ and then starts decreasing after reaching a certain value of $n$. Where this maximum occurs depends on other parameters.

\subsection{Asymptotic behavior for the case of circular orbits}\label{sec:asympt-beh}

We now move to the discussion of the asymptotic behaviour of the power loss in two limiting cases $m_M \ll \Omega$ and $m_M \gg \Omega$, where $m_M$ is the mass of the mediator, $M = A,\, \phi$. In this subsection, for simplicity we consider the straightforward case of circular orbits ($e=0$) and massless fermions ($m_\psi = 0$). For the eccentricity dependent part of the power loss, $K(n,e)$, we have
\beq 
\lim_{e\, \to\, 0} K(n,e) = 
\lim_{e\, \to\, 0} \left(J^\prime_n(ne)^2 + \frac{1 - e^2}{e^2}J_n(ne)^2\right) = {1 \over 2} \delta_{n,1}. 
\eeq
Thus the only mode that contributes to the power loss in the circular orbit limit is the mode with $n=1$.

First, let us consider the regime of light mediators, $m_M \ll \Omega$, or equivalently $n_M \ll 1$. In this limit, $F^M(x)$ defined in Eq.~\eqref{eq:FM-gen-mai} is dominated by the second term. We thus neglect the first and the third terms of $F^M(x)$ and take the second term's limit $n_M \to 0$. After that, the integral in \eqref{eq:Bn-mai} can be performed analytically, yielding the following asymptotic expressions for the power radiated via vector and scalar, respectively: 
\begin{equation}\label{eq:PA-small-m}
    P^A(m_A \ll \Omega) \approx \frac{g^2}{6\pi}   Q^2 a^2 \Omega^4,
\end{equation}
\begin{equation}
    P^\phi(m_\phi \ll \Omega) \approx \frac{g^2}{12\pi} N^2 a^2 \Omega^4.
\end{equation}
The asymptotic behavior that we find for $P^A$ and $P^\phi$ reproduces the known results for the on-shell vector~\cite{KumarPoddar:2019ceq, Dror:2019uea,Krause:1994ar} and scalar~\cite{Krause:1994ar} radiation. This is expected as, in the regime $m_M \ll \Omega$, the fermion pair radiation is dominated by on-shell boson production. Additionally, Eq.~\eqref{eq:PA-small-m} also reproduces the Larmor formula for the power of the electromagnetic wave radiation given in Eq.~\eqref{eq:larmor}. To see this, recall that the acceleration on a circular orbit is equal to $a\Omega^2$, where $a$ is the radius of the orbit and $\Omega$ is the frequency of revolution.

Next, we study the regime when on-shell boson production is kinematically forbidden, and the fermion pair radiation takes place through the off-shell mediator. This is the limit of heavy mediators, $m_M \gg \Omega$, or equivalently $n_M \gg 1$. As in the case of the light mediators, we take the $n_M \to \infty$ limit of $F^M(x)$ and find that the resulting expression can be integrated analytically. Upon performing the integration, we find that the vector and scalar-mediated radiation behave as 
\begin{equation}\label{eq:PA-heavy-m}
P^{A}(m_A \gg \Omega) \approx \frac{g^4 q_\psi^2 Q^2}{210 \pi^3} \frac{a^2 \Omega^8}{m_A^4} = \frac{1}{35\pi^2} \frac{g^2 q_\psi^2\Omega^4}{m_A^4} \times P^{A}(m_A \ll \Omega),
\end{equation}
\begin{equation}\label{eq:Pphi-heavy-m}
P^{\phi}(m_\phi \gg \Omega) \approx \frac{g^2 g^{\prime2}N^2}{840 \pi^3}\frac{a^2 \Omega^8}{m_\phi^4} = \frac{1}{70\pi^2} \frac{g^{\prime 2}\Omega^4}{m_\phi^4} \times P^{\phi}(m_\phi \ll \Omega).
\end{equation}
We learn that in the limit of heavy mediators, the fermion pair radiation is suppressed compared to on-shell boson radiation by the following factors:
\begin{enumerate}
\item 
A factor of $g^2 q_\psi^2$ or $g^{\prime 2}$, which, at the amplitude level, comes from the coupling of the mediator to the fermion pair. 
\item
A factor of $\Omega^4/m_\phi^4$, which comes from the propagator of the mediator.
\item
A phase space factor of $1/35\pi^2$ or $1/70\pi^2$, which arises from the fact that there are more particles in the final state in the case of the off-shell pair production than in the case of the on-shell boson production.
\end{enumerate}

Note that Eqs.~(\ref{eq:PA-heavy-m}) and (\ref{eq:Pphi-heavy-m}) can be interpreted as integrating out the heavy mediator, resulting in an effective 4-Fermi interaction with a coefficient proportional to $g^2/m_A^2$ or $gg'/m_\phi^2$. Thus, it is also valid for $t$-channel and $u$-channel interactions.

Last, we compare the results of the vector to that of the scalar mediators. Consider $m_A = m_\phi$, $Q^2 = N^2$ and $g' = g q_\psi$. In this case, the power radiated via the vector mediator is greater than the power radiated via the scalar mediator in both radiation regimes. In particular, we have
\begin{equation}
    \frac{P^A (m_A \ll \Omega)}{P^\phi (m_\phi \ll \Omega)} \approx 2, \qquad \frac{P^A (m_A \gg \Omega)}{P^\phi (m_\phi \gg \Omega)} \approx 4.
\end{equation}
These factors are related to the different number of degrees of freedom between the vector and scalar cases. There are two polarization states for an on-shell massless vector, while the scalar has only one. For the deeply off-shell mediator, the correspondence is not so clear, but it seems to us that it is related to the fact that off shell gauge boson, $A_\mu$, has four degrees of freedom

\subsection{Fermion-pair radiation in the SM }\label{sec:fermion-pair-rad-SM}

The expression in Eq.~\eqref{eq:PA-heavy-m} can be used to estimate the power loss due to fermion pair radiation by classical sources within the SM. In this subsection, we consider neutrino pair radiation mediated by $Z$-boson. The contribution due to $W$-boson mediated pair emission is qualitatively the same as the $Z$-boson contribution and is expected to be of the same order. The main difference between the two contributions is due to the fact that $W$-boson mediated radiation is only relevant for leptons in the source while $Z$-boson contribution is present for all types of fermions.

Consider a source made of $N_\Psi$ fermions of type $\Psi$ with the total weak charge $Q = N_\Psi q_\Psi$. To apply Eq.~\eqref{eq:PA-heavy-m} to the neutrino pair radiation in the SM, we need to recall that Eq.~\eqref{eq:PA-heavy-m} was derived under the assumption of vectorial couplings, while the SM is a chiral theory. The relevant parts of the SM Lagrangian are different from the Lagrangian in Eq.~\eqref{eq:L-vec-int}; in particular, in the SM we have
\begin{equation}
    \mathcal{L}_{\text{SM}} \supset -i\frac{g}{2\cos \theta_W}\left( \bar\Psi \gamma^\mu (c_V^\Psi - c_A^\psi)\Psi + \bar\nu \gamma^\mu (c_V^\nu - c_A^\nu)\nu\right)Z_\mu.
\end{equation}
Thus Eq.~\eqref{eq:PA-heavy-m} yields the following expression for the $Z$-boson mediated power loss due to the neutrino pair radiation in the SM
\begin{equation}\label{eq:PZ-1}
    P^Z(m_Z \gg \Omega) \approx \frac{1}{210 \pi^3}\frac{g^4 q_\nu^2 q_\Psi^2 N_\Psi^2}{16\cos^4 \theta_W} \frac{a^2 \Omega^8}{m_Z^4},
\end{equation}
where we perform the replacement $g \rightarrow g/(2\cos \theta_W)$ in Eq.~\eqref{eq:PA-heavy-m} and define
\begin{equation}
    q_\psi^2 = q_\nu^2 =  (c_V^{\nu})^2 + (c_A^\nu)^2, \qquad q_\Psi = c_V^\Psi, \qquad m_A = m_Z\,.
\end{equation}
Note that, for the source, only vectorial coupling $c_V^\Psi$ enters the power loss. This is because we consider coherent radiation.

The expression in Eq.~\eqref{eq:PZ-1} can be rewritten as
\begin{equation}\label{eq:Ploss-Z-GF}
    P^Z (m_Z \gg \Omega) \approx G_\text{eff}^2 q_\Psi^2 q_\nu^2 N_\Psi^2 \frac{a^2\Omega^8}{210 \pi^3}\,,
\end{equation}
where $G_\text{eff} = \sqrt{2}G_F$ and $G_F$ is the Fermi constant. When the power loss is written in the form of Eq.~\eqref{eq:Ploss-Z-GF}, it becomes clear that it is the same as what one would obtain by performing the calculation for the effective Fermi theory with the effective Lagrangian given by
\beq
\mathcal{L}_\text{eff}^Z \supset G_{\text{eff}}
[\bar{\Psi}\gmu 
(c_V^\Psi- c_A^\Psi\gfive)
\Psi][\bar{\nu}\gMu
(c_V^\nu- c_A^\nu\gfive)\nu].\label{fourfermi}
\eeq
This, of course, is not surprising as we consider radiation at the energy $\Omega$, which is much less than the electroweak scale, $\Omega\ll m_Z$. In fact, the result in Eq.~\eqref{eq:Ploss-Z-GF} applies to any effective 4-Fermi interaction. While we derive our results for $s$-channel exchange, in the limit where the mediator is much heavier than the orbit frequency, we do not need to distinguish between $s$-channel and $t$-channel. Thus, Eqs.~\eqref{eq:PZ-1} and~\eqref{eq:Ploss-Z-GF} can also be used for $t$-channel $W$-exchange in the SM.

Finally, we discuss the situation when there are several different types of fermions in the source. In this case, we need to first add all the amplitudes that correspond to the radiation by different fermions $\Psi$ (for leptons, we add both $Z$-boson and $W$-boson contributions). Then, we square the sum of the relevant amplitudes to obtain the total emission rate.

We end this subsection with the following remark.
The power loss due to neutrino pair radiation in the SM was estimated in Ref.~\cite{Mohanty:1994yi} to be $P^Z_{SM} \sim G_F^2 \Omega^6$.
Using the explicit calculation, however, we find that $P^Z_{SM} \sim G_F^2 a^2 \Omega^8$. That is, there is an extra factor of $a^2\Omega^2$ compared to the estimation of Ref.~\cite{Mohanty:1994yi}. In fact, our result includes the semi-major axis $a$ as an additional energy scale of the system. 

\section{Fermion pair radiation by pulsar binaries} \label{sec:Rad_NS}

We now move to discuss the phenomenological applications of our results to astrophysical systems. We focus on the neutrino-pair emission from pulsar binaries ~\cite{vanStraten:2001zk,Kramer:2006nb,Stairs:2002cw,Shannon:2013dpa,Antoniadis:2013pzd,Bhat:2008ck,Freire:2012mg,Ferdman:2014rna,vanLeeuwen:2014sca,Jacoby:2006dy,Davoudiasl:2019nlo}. A pulsar binary is a binary system of a pulsar and companion.
This choice is motivated by the availability of extensive period decay data for such systems. In particular, we apply our results to two binaries: Hulse-Taylor binary PSR B1913+16~\cite{Hulse:1974eb,Taylor:1982zz,weisberg2016relativistic} (a system of a pulsar and a neutron star) and PSR J1738+0333~\cite{Freire:2012mg,kilic2015psr} (a system of a pulsar and a white dwarf). The parameters characterizing the two systems are summarized in Table~\ref{tab:starstats}. 

In what follows, we first discuss the applicability of our results of Section~\ref{sec:plossformulae} to pulsar binaries in general. Then we estimate the contribution to the power loss due to neutrino pair emission in the SM and show that it is negligible compared to the gravitational wave radiation. We then consider neutrino pair radiation in two BSM scenarios via ultralight vector and scalar mediators and apply our results to the pulsar binaries with the parameters in Table~\ref{tab:starstats}.

\begin{table}[]
\begin{tabular}{|c|c|c|}
\hline    
\textbf{Binary system} & \textbf{~~PSR B1913+16~\cite{weisberg2016relativistic}~~} & \textbf{~~PSR J1738+0333~\cite{Freire:2012mg}~~}  \\
\hline
Eccentricity $e$ & 0.6171340(4) & $3.4(11) \times 10^{-7}$\\
\hline
Pulsar mass $m_1$ ($M_\odot$) & $1.438(1)$ & $1.46(6)$\\
\hline
Companion mass $m_2$ ($M_\odot$) & $ 1.390(1)$ & $0.181(8)$\\
\hline
Binary period $T_b$ ($\text{GeV}^{-1}$)& $4.240\times10^{28}$ & $4.657\times 10^{28}$\\
\hline
Intrinsic period decay $\dot{T}_b$ &$-2.398(4)\times 10^{-12}$ & $-2.59(32) \times 10^{-14}$\\
\hline
~~Predicted period decay due to GW $\dot{T}_{GW}$~~ & $-2.40263(5)\times 10^{-12}$ & $-2.77(19)\times 10^{-14}$\\
\hline
Ratio of period decays $\mathcal{R} = \dot{T}_b/\dot{T}_{GW}$ & $0.9983(16)$ & $0.94(13)$\\
\hline
\hline
Orbital frequency $\Omega = 2\pi/T_b$ ($ \text{GeV}$)& $1.482 \times 10^{-28}$ & $1.349 \times 10^{-28}$\\
\hline
~~Semi-major axis $a$ ($ \text{GeV}^{-1}$) ~~& $9.878 \times 10^{24}$ & $8.77 \times 10^{24}$\\
\hline
\end{tabular}
\caption{The relevant parameters for the
PSR B1913+16 and PSR J1738+0333 binary systems. Figures in parenthesis are the $1\sigma$ uncertainties in the last quoted digit, where all the uncertainties are symmetrized.  $M_{\odot}
$ is the mass of the sun. The relative experimental error of the binary period $T_b$ is $\sim10^{-12}$ for PSR B1913+16, and $\sim10^{-11}$ for PSR J1738+0333. The double line separates binary parameters quoted in Ref.~\cite{weisberg2016relativistic, Freire:2012mg} and the ones we derive.
Values of the semi-major axis $a$ are calculated using Eq.~\eqref{eq:Omega-def}.
}
\label{tab:starstats}
\end{table}
%

\subsection{Pulsar binaries as a classical source}

The results for the fermion pair radiation, summarized in Eqs.~\eqref{eq:Ploss-A-mai}-\eqref{eq:f0f1f2scal-mai}, were derived for the case of classical current describing non-relativistic point-like object following an elliptical orbit. 
To justify the application of our results to pulsar binaries, we note the following:
\begin{enumerate}
\item
A pulsar binary can be treated as a classical source. The typical size of a pulsar binary can be estimated as the size of the semi-major axis which varies between $10^{6}$ and $10^{8}$ km, that is, $a  \sim 10^{24} - 10^{26} \text{ GeV}^{-1}$. The wavelength of the radiation is determined by the fundamental frequency of the orbit, and for a typical pulsar binary with periods in the range of $10^{-1} - 10^{3}$ days, the wavelength is $\lambda \sim 10^{28} - 10^{32} \text{ GeV}^{-1}$. Thus, $\lambda \gg a$ and we conclude that pulsar binaries can be treated as classical radiation sources.

\item 
Stars of the pulsar binary can be treated as point-like objects. Typical sizes of stars in a binary vary from $r \sim 10\,{\text{km}} \sim 10^{19} \text{ GeV}^{-1}$, for neutron stars, and $r \sim 10^3 \,{\text{km}} \sim 10^{21} \text{ GeV}^{-1}$, for white dwarfs. Thus $r \ll a,\, \lambda$ and both pulsar and its companion can be treated as point-like objects. Moreover, $r \ll \lambda$ implies the coherence of the radiation.
\item 
The motion of the pulsar and its companion in the binary system is non-relativistic. We can roughly estimate the orbital velocity of the stars in a binary as $v \sim a\Omega$, which for characteristic values quoted above implies $v \lesssim 10^{-2}$.
\item 
For a wide range of pulsar binary systems, the observed power loss is such that it has no significant effect on the eccentricity of the orbit. Thus we can treat the orbit as elliptical over the time of observation. For example, the Hulse-Taylor binary has $e\sim1$, with $T_b (\D e/ \D t) \lesssim 10^{-11}$, where $T_b$ is the binary period and $\D e/\D t$ is the time derivative of the eccentricity~\cite{weisberg2016relativistic}.
\end{enumerate}

Now that we have established that the results of Section~\ref{sec:plossformulae} can be applied to pulsar binaries, we proceed in two steps. First, we modify our expressions for the classical current and number density in Eqs.~\eqref{eq:jcldef-ellipse} and~\eqref{eq:rhocldef-ellipse} to the case of two point-like objects on an elliptical orbit. Second, we perform the standard reduction of the two-body problem to a one-body problem.

We write the classical current and number density as
\beq 
J_{\text{cl}} ^{\mu} (x) = \sum_{b = 1,2} Q_b\, \delta^3 ({\bm x} - {\bm x}_b(t) ) u_b ^ {\mu}, \label{eq:jcldef}
\eeq
and
\beq 
\rho_{\text{cl}} (x) = \sum_{b = 1,2}N_b\, \delta^3 ({\bm x} - {\bm x}_b(t) \, , \label{eq:rhocldef}
\eeq
respectively. Here, $b=1,2$ is the index that labels the stars of the binary system, ${\bm x}_b (t)$ is the position of the $b$-th star at time $t$, and $u_b^{\mu}$ is its four-velocity.

Next, we move to the binary system's Center-of-Mass (CoM) frame. For that, we define $\bm{R}$, the coordinate of center of mass, and $\bm{r}$, the distance between the two stars,
\begin{eqnarray}
    \bm{R} = \frac{m_1}{m_1+m_2} \bm{x}_1 + \frac{m_2}{m_1+m_2}\bm{x}_2, \qquad
    \bm{r} = \bm{x}_1 - \bm{x}_2\,,
\end{eqnarray}
where $m_1$ and $m_2$ are the masses of the two stars. 


As we are not concerned with the translational motion of the system as a whole, which is described by $\bm{R}$, we can solely focus on $\bm{r}$. This is the standard two-body to one-body problem reduction for central force motion. The non-relativistic classical trajectory of the stars in the CoM frame can thus be described by the vector $ \bm{r} = (x, y, 0)$ and is given by elliptical orbits as in Eq.~\eqref{eq:umu}:
\beq x = a(\cos \xi -e) , \ \ y = a \sqrt{1-e^2} \sin \xi , \ \ \ \Omega t = \xi - e \sin \xi,
\eeq
where $e$ is the eccentricity, $a$ is the semi-major axis of the elliptical orbit, and the fundamental frequency of revolution is given by
\begin{equation}\label{eq:Omega-def}
    \Omega= \sqrt{\frac{G_N(m_1+m_2)}{a^{3}}}\,.
\end{equation}

The results of Eqs.~\eqref{eq:Ploss-A-mai}-\eqref{eq:f0f1f2scal-mai} generalize to the case of binary systems via the following replacements that follow from the 2-body to 1-body reduction procedure:
\beq
Q^2 \to M^2 \left({Q_1 \over m_1} - {Q_2 \over m_2} \right)^2, \qquad
N^2 \to M^2 \left({N_1 \over m_1} - {N_2 \over m_2}\right)^2, \label{eq:2bodyreplace}
\eeq
where
\beq
M = {m_1 m_2 \over  m_1 +m_2} \label{eq:redmass}
\eeq
is the reduced mass of the binary system. As a result we obtain the following expressions for the power loss in $n$-th harmonic for a vector and scalar mediators respectively:
\begin{align}
P^{A}_n &= {g^4 q_{\psi}^2 \over 12 \pi^3}  M^2 \left ( {Q_1 \over m_1} - {Q_2 \over m_2} \right)^2 a^2 \Omega^4\, 
 B_n^A(n_A,n_\psi,n_\Gamma), \label{eq:Ploss2-A-mai} \\
P^{\phi}_n &= \frac{g^2 g'^2}{12 \pi^3}M^2 \left ( {N_1 \over m_1} - {N_2 \over m_2} \right)^2 a^2 \Omega^4\,
B_n^\phi(n_\phi,n_\psi,n_\Gamma),\label{eq:Ploss2-phi-mai}
\end{align}
where the functions $B_n^A$ and $B_n^\phi$ are defined in Eqs.~\eqref{eq:Bn-mai}-\eqref{eq:f0f1f2scal-mai}.


\subsection{Neutrino pair radiation by pulsar binaries in the SM}

In the SM, for the pulsar binary, the power loss via electroweak mediators is discussed in Sec.~\ref{sec:fermion-pair-rad-SM}.
Here, we simply generalize it to the case of 2-body motion using Eq.~\eqref{eq:2bodyreplace}. We obtain the following expression for the power loss in neutrino pair radiation via $Z$-exchange in the SM
\begin{equation}
    P_{\text{SM}} \approx \frac{G_F^2 \left({c_V^{\nu}}^2 + {c_A^\nu}^2\right)}{105 \pi^3 \cos^2 \theta_W} M^2a^2 \Omega^8 \left( \frac{1}{m_1}\sum_{i = n, p, e, \dots} c_V^i N_{1i}Q_{1i}- \frac{1}{m_2}\sum_{i = n, p, e, \dots} c_V^i N_{2i}Q_{2i}\right)^2
\end{equation}
where the sum goes over all microscopic constituents of binary stars, such as neutrons ($n$), protons ($p$), electrons ($e$), etc. To perform a numerical estimate, we consider a pulsar binary with a neutron star companion and assume that all of the neutron star mass is in the form of neutrons. We consider a typical pulsar-neutron star binary with 
\beq
m_{1,2} \sim M_{\odot} \sim 10^{57} \text{GeV}, \qquad
a \sim 10^{25}\;\text{GeV}^{-1}, \qquad \Omega \sim 10^{-28} \  \text{GeV},
\eeq
and non-zero dipole moment
\begin{equation}
    M^2\left(\frac{Q_1}{m_1} - \frac{Q_2}{m_2}\right)^2 \sim Q_{1,2}^2 \sim 10^{114},
\end{equation}
where $Q_{b} = N_{b}(n) - N_{b}(\bar{n}) \approx N_{b}(n)\approx M_\odot/m_n \approx 10^{57}$, with $b=1,2$, are the neutron charges of the neutron stars, $N_b(n)$ and  $N_b(\bar{n})$ are the numbers of neutrons and anti-neutrons respectively, $m_n$ is the neutron mass. Using $c_V^\nu = c_A^\nu = 1/2$, $c_V^n = -1/2$, and the measured values of $m_n$, $G_F$, and $\theta_W$,
we find the following numerical estimate for the radiated power
\beq \label{eq:SM-estimate-num}
P_{\text{SM}} \sim 10^{-56}\text{eV}^2
.
\eeq

To see if the above result is significant, we compare it to the power loss in the form of gravitational wave (GW) radiation. Using the quadrupole formula for the GW radiation~\cite{Peters:1963ux} for the case of circular orbit ($e = 0$) we have
\begin{equation}\label{eq:P-GW-est}
P_\text{GW} = \frac{32}{5}G_N M^2 a^4\Omega^6 \sim 10^8\, \text{GeV}^2 
\end{equation}
where $G_N$ is Newton's gravitational constant. The rough estimates in Eqs.~\eqref{eq:SM-estimate-num} and~\eqref{eq:P-GW-est} show that, in the SM, the fermion-pair radiation by astrophysical objects is completely negligible compared to the gravitational wave radiation. 

We close the subsection with one remark. Within the SM,
neutron stars also emit synchrotron radiation of fermion-antifermion pairs in their self-produced magnetic fields, as shown in Ref.~\cite{Yakovlev:2000jp}. This phenomenon is different from the one we consider here. Synchrotron radiation is an incoherent effect. Thus, the power loss, in this case, scales as $N$, the number of neutrons in the star. In the case we are considering, the radiation is coherent and comes from the star's acceleration as a whole. Then, the net power that is radiated is proportional to $N^2$.


\subsection{New physics constraints from the neutrino pair radiation by pulsar binaries}
\label{sec:nsappl}

Since extra radiation in the SM is negligible, 
any observed deviation from the gravitational wave radiation would be strong evidence for the physics beyond the SM. In particular, fermion-pair radiation can be enhanced in BSM models with light vector or scalar mediators, with $m_{A, \phi} \ll m_Z$. To explain why such light bosonic states have evaded detection so far, we must require that they have small couplings, thus  evading all the available constraints. The smallness of couplings, however, still can be compensated in cases where the object has a large charge under the new symmetries. This can be the case for astrophysical objects. Thus, such objects are our prime focus in the rest of this work.

In particular, in this subsection, we demonstrate how our results can be used to derive new physics bounds from the neutrino pair radiation by pulsar binaries. As we mentioned above, we use two distinct pulsar binary systems, the Hulse-Taylor binary PSR B1913+16 and PSR J1738+0333. The relevant properties of the two systems are summarized in Table.~\ref{tab:starstats}. The Hulse-Taylor binary is a pulsar binary with a neutron star companion, it is highly eccentric, and  the mass ratio of the two stars is close to 1. The PSR J1738+0333, on the other hand, is a pulsar-white dwarf binary with an almost circular orbit and a high pulsar-to-companion mass ratio. For both systems, the data on the orbital period decay is shown in  Table~\ref{tab:starstats}. Both binaries lie within $1\sigma$ of the general relativity prediction.

In our analysis, we exploit the fact that typical neutron stars contain a very large number of muons, $N(\mu) \sim10^{55}$~\cite{Garani:2019fpa, Pearson:2018tkr,Harry:2018hke,Zhang:2000my}. Thus, the effects of muonophilic new physics can be significantly enhanced. The presence of the large muon number in neutron stars is attributed to the fact that when the electron chemical potential, $\mu_e$, is larger than the muon mass $\mu_e > m_\mu$, it becomes energetically favorable for relativistic electrons at the Fermi surface to decay into muons via $e^- \to \mu^- + \bar{\nu}_{\mu} + \nu_e$. Moreover, the muonic beta-decay $n\to p + \mu^-  + \bar{\nu}_{\mu}$ and inverse beta-decay $p + \mu^- \to n + \nu_\mu$ reactions become energetically favorable, while the muon decay $\mu^- \rightarrow e^- + \bar\nu_e + \nu_\mu$ is forbidden by Fermi statistics.

Being motivated by the neutron star muonic content, we consider neutrino pair emission by pulsar binaries via the following two types of BSM mediators:
\begin{itemize}
\item $\operatorname{U}(1)_{L_\mu - L_\tau}$ massive gauge boson with
\begin{equation}\label{eq:L-gauge-int}
\mathcal{L} \supset g A_\alpha \left(\bar \mu \gamma^\alpha \mu - \bar \tau \gamma^{\alpha} \tau +   \bar \nu_\mu \gamma^\alpha \nu_\mu -   \bar \nu_\tau \gamma^\alpha \nu_{\tau}\right),
\end{equation}
\item Massive muonophilic scalar with
\begin{equation}\label{eq:L-scalar-int}
\mathcal{L} \supset g \phi \bar \mu \mu +  g^\prime \phi \bar \nu_\mu \nu_\mu\,.
\end{equation}
\end{itemize}

It is known that at least two of the SM neutrinos are massive, while the third neutrino can be very light or massless. This means that only one neutrino mass eigenstate can be radiated in the two scenarios we consider here.  A realistic treatment of neutrino emission would include insertions of the corresponding PMNS matrix elements~\cite{Maki:1962mu}, resulting in an additional factor of order one. Since we already neglecting an ${\cal O}(1)$ factor coming from the estimate of the muon number density in the neutron stars, we also ignore any PMNS factors in the rest of this section. 

Note also that in a theory with general couplings to the left and right-handed neutrinos, i.e., $g A_{\alpha} \bar{\nu} \gamma^\alpha (c_V -c_A \gamma^5) \nu$, the results for the power loss are qualitatively similar. Moreover, in the case of massless neutrinos, the power loss for the case of the general coupling is the same as the power loss for the case of purely vectorial coupling up to $g^2 \rightarrow g^2 (c_A^2 + c_V^2)$ replacement. This is why in what follows, for simplicity, we consider the case of the vectorial coupling only.

These two BSM models imply the possibility for the neutrino pair radiation at rates enhanced compared to the SM. Our results from Eqs.~(\ref{eq:Ploss2-A-mai}) and (\ref{eq:Ploss2-phi-mai}) thus can be used to set bounds on the coupling constants and masses of the new bosons.

The presence of the muonophilic new physics, however, not only alters the radiation patterns of pulsar binaries, but it also has important implications for the neutron star's equation of state. In particular, the presence of a repulsive (vector) or attractive (scalar) interaction between muons could affect the muon number, which depends on the coupling $g$ to the new physics. In the following, we write the muon number as $N(\mu,g)$ to keep the dependence on $g$ explicit. 

The number of muons becomes $g$-dependent as the interactions change the muon chemical potential. The muon interaction due to the $L_\mu-L_\tau$ vector boson is repulsive, and thus the chemical potential is increased compared to its SM value by $\varepsilon \sim g^2 N(\mu, g)/R$, where $R$ is the radius of the neutron star the boson mass is neglected. When the coupling $g$ is small, such that $\varepsilon \ll m_\mu$, the effect of the new interaction is insignificant, and the number of muons is approximately given by its value in the limit of no interaction $N(\mu, g=0)$. When the interaction is strong, such that $\varepsilon \gg m_\mu$, it becomes energetically less favorable to have muons inside the neutron star and thus $N(\mu, g) < N(\mu, g=0)$. 

Similar reasoning applies to the case of the scalar mediator. The only difference is the sign of the interaction. In the scalar case, the interaction between muons is attractive. Thus the muon chemical potential is decreased by $\varepsilon$. This leads to the increase of the muon number for larger couplings $N(\mu,g) > N(\mu,g=0)$. In both cases, the change from the regime when $N(\mu,g) \approx N(\mu,g=0)$ to the situation when the interaction starts to affect the muon number happens for couplings such that $\varepsilon \sim m_\mu$, or numerically $g\sim 10^{-18}$ for a typical neutron star~\cite{Dror:2019uea}.

However, in what follows, we ignore the effect of the new physics on the muon number. Everywhere in our analysis, we use the muon number in the limit of no new physics interaction, that is we set $N(\mu) = N(\mu, g=0)\sim 10^{55}$~\cite{Garani:2019fpa, Pearson:2018tkr,Harry:2018hke,Zhang:2000my}. In principle, $g$-independence of muon number can be achieved in models with both vector and scalar mediators with fine-tuned coupling constants such that the repulsive and attractive interactions cancel each other.

To apply Eqs.~(\ref{eq:Ploss2-A-mai}) and (\ref{eq:Ploss2-phi-mai}), we define $N_b (\mu)$ and $N_b (\bar{\mu})$ as the number of muons and antimuons respectively in neutron star labeled by $b = 1,2$. Then, as there are almost no tau leptons in neutron stars, $Q_b = N_b (\mu) - N_b (\bar\mu)$ is the total charge of the neutron star under the $L_{\mu}-L_{\tau}$ gauge symmetry, and $N_b = N_b(\mu) + N_b (\bar\mu)$ is the total number of muons and anti-muons in the star. Additionally, since $N_b(\bar{\mu}) \approx 0$, we have $Q_b \approx N_b$.

The energy lost through radiation in a binary star system can be directly probed by measuring the decay of the orbital period. Assuming that the attractive gravitational force between the two stars is such that their orbits stay Keplerian, the decay rate of the period of revolution $T_b$ is related directly to the energy lost via radiation~\cite{KumarPoddar:2019ceq}:
\beq
\dot{T}_b = -6 \pi a^{5/2}  G_N^{-3/2} (m_1 m_2)^{-1} (m_1+m_2)^{-1/2} \times P_{\text{loss}}, \label{eq:dpbdt}
\eeq
where $\dot{T}_b$ is the time derivative of the binary period, $G_N$ is the gravitational constant, $m_1$ and $m_2$ are the masses of the stars in the binary system, $a$ is the semi-major axis of the elliptical orbit, and $P_\text{loss}$ is the total power radiated. The decay of the period per unit of time is dimensionless and is measured experimentally.

GW emission is the dominant source of power loss in a binary star system. Assuming that the GW emission and neutrino pair emission are the only sources of energy loss, we have
\beq P_{\text{loss}} = P_{\text{GW}} + P_{\bar{\nu}\nu}, \eeq
where $P_{\bar{\nu}\nu}$ is the power loss due to the neutrino pair radiation and $P_{GW}$ is the power loss due to GW emission, which, to the leading order, is given by the GW quadrupole radiation formula~\cite{Peters:1963ux},
\beq
P_{\text{loss}}^{GW} = {32 \over 5} G \Omega^6 M^2 a^4 (1-e^2)^{-7/2} \left(1+ {73 \over 24} e^2 + {37 \over 96} e^4 \right) \label{eq:powgw}, \eeq
where $M$ is the reduced mass of the system, as defined in Eq.~\eqref{eq:redmass}. The binary period decay $\dot{T}_b$ thus can be written as a sum of two contributions,
\begin{equation}
    \dot{T}_b = \dot{T}_\text{GW} + \dot{T}_{\bar \nu \nu}\,.
\end{equation}

We next introduce the period decay ratio $\mathcal{R}$ as the ratio of the measured period decay to the theoretical prediction of the period decay due to GW radiation,
\begin{equation}
    \mathcal{R} = \frac{\dot{T}_b}{\dot{T}_\text{GW}} = 1 + \frac{\dot{T}_{\bar \nu \nu}}{\dot{T}_\text{GW}}.
\end{equation}
We use the measured value of $\mathcal{R}$ to set $2\sigma$ limits on the masses and couplings of the BSM mediators of neutrino pair radiation as
\begin{equation}\label{eq:bound}
    \frac{\dot{T}_{\bar \nu \nu}}{\dot{T}_\text{GW}} \leq \left(\mathcal{R} - 1\right) + 2\sigma\,.
\end{equation}

\begin{figure}[t]
\centering
\includegraphics[width=0.49\textwidth]{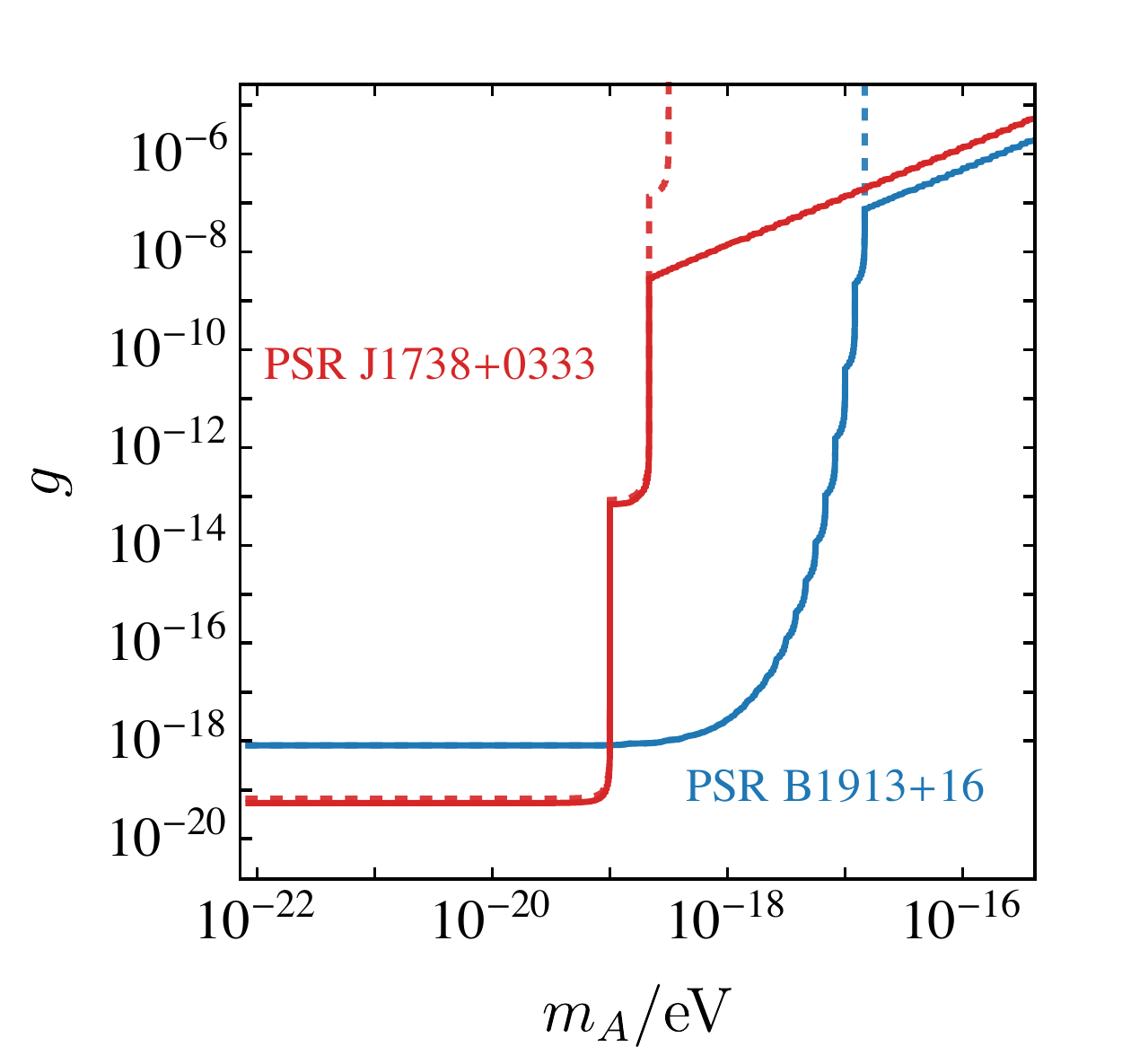} 
\includegraphics[width=0.49\textwidth]{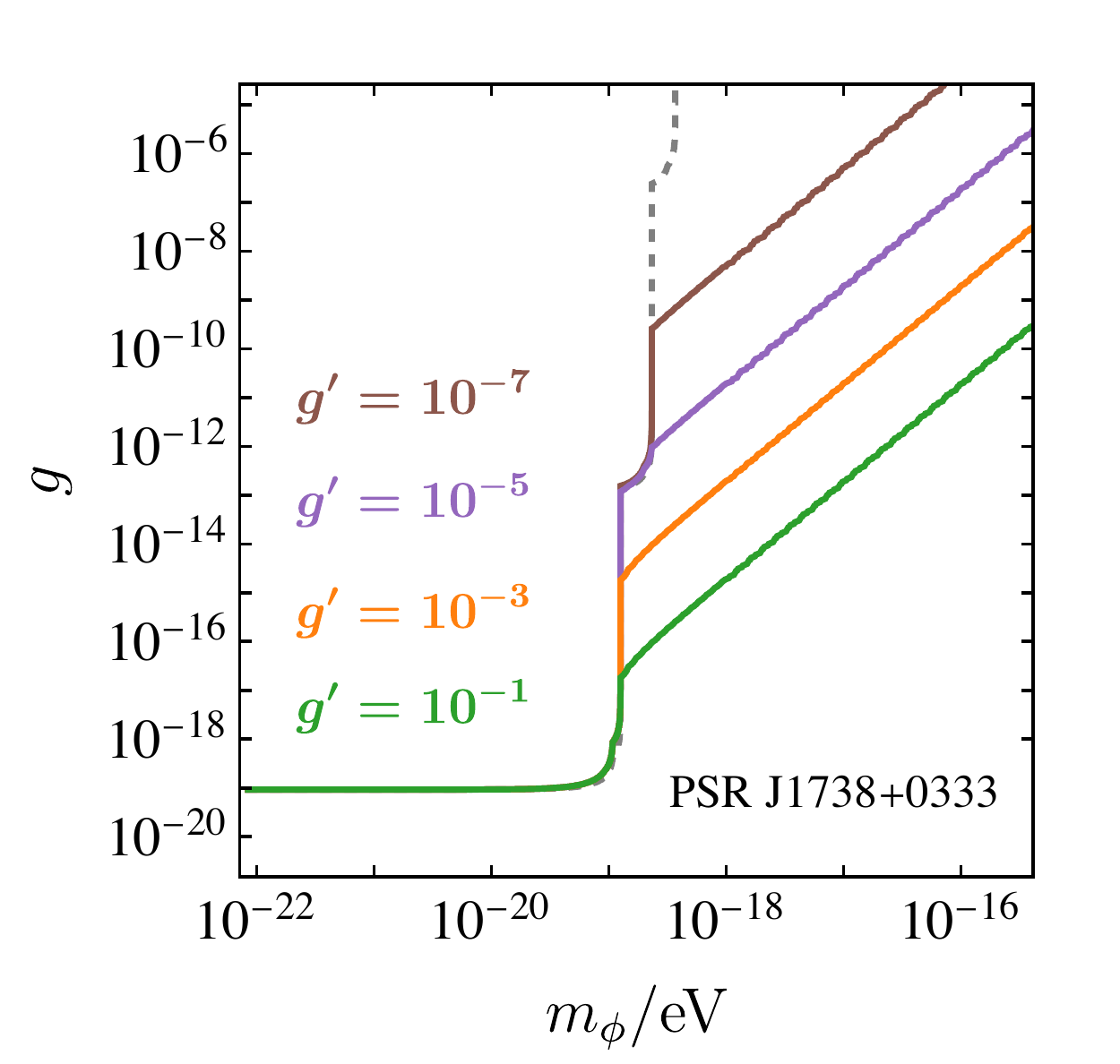}
\caption{{\it Left:} 
Constraints on $g \,{\text{vs}}\, m_{A}$ from the highly eccentric PSR B1913+16 (Hulse-Taylor)
Bounds from the neutrino pair radiation (solid) and vector boson radiation (dashed) are shown such that the region above the curves is excluded by the measurements of the period decay. The system parameters are taken from Table~\ref{tab:starstats}. {\it Right:}  Constraints on $g \,{\rm vs}\, m_{\phi}$ from PSR J1738+033. The dashed gray line corresponds to the bound set by the emission of the scalar boson only, while the solid lines show the bounds from including a coupling $g'$ to the neutrinos.}
\label{fig:exclplots}
\end{figure}

The resulting constraints on the parameter space $(g, m_A)$ and $(g, m_\phi)$ that we derive from the period decay data for the Hulse-Taylor binary and PSR J1738+033 are shown in Fig.~\ref{fig:exclplots}. When deriving the constraints, we use $Q_b = N_b = 10^{55}$ with $b=1,2$ and $q_\nu = 1$. For the gauge boson mediator (left panel), we calculate the period decay due to neutrino pair emission, $\dot{T}_{\bar \nu \nu}$, using Eqs.~\eqref{eq:Ploss2-A-mai} and~\eqref{eq:dpbdt}. As we take all three neutrinos to be massless, and as $L_\mu-L_\tau$ boson couples to two neutrino types, there is an extra factor of $2$ in Eq.~\eqref{eq:Ploss2-A-mai}. Similarly, for the case of the scalar mediator (right panel), we use Eqs.~\eqref{eq:Ploss2-phi-mai} and~\eqref{eq:dpbdt}. As there is no symmetry that requires equality of $g$ and $g^\prime$ in the case of the scalar mediator, we present our results for the scalar case in the $(g, m_\phi)$ plane for four different values of $g^\prime$ that vary from $10^{-7}$ to $10^{-1}$. 

First, let us discuss the left panel of Fig.~\ref{fig:exclplots}, which shows constraints on the mass and coupling of the gauge boson. For the PSR J1738+0333 (red line), whose orbit is very close to circular, the effect of neutrino pair radiation becomes significant for the mediator masses greater than the second harmonic frequency, $m_A > 2 \Omega$. For the highly eccentric Hulse-Taylor binary, off-shell radiation dominates for $m_A > 85\Omega$. In the region $m_A>2\Omega$ ($m_A > 85\Omega$) for PSR J1738+0333 (Hulse-Taylor binary), the boundary of the excluded region is approximately quadratic in the mediator mass. 
This is in stark contrast with the case of the on-shell boson emission discussed in Ref.~\cite{KumarPoddar:2019ceq, Dror:2019uea, Krause:1994ar}, where the boundary of the excluded region jumps in steps at $m_{A} = n\Omega$, with $n$ being an integer. For comparison, the dashed lines in Fig.~\ref{fig:exclplots} show the bounds due to the on-shell boson radiation. 

Finally, we comment on the right panel of Fig.~\ref{fig:exclplots}, which shows the constraints on the mass $m_\phi$ and coupling $g$ for different values of $g^\prime$ in the case of the scalar mediated radiation. We only demonstrate the constraints for PSR J1738+0333; the results for the Hulse-Taylor binary are qualitatively the same. Depending on the value of $g^\prime$ the off-shell scalar radiation starts to dominate for $m_\phi > \Omega$ ($g^\prime \gtrsim 10^{-4}$) or $m_\phi > 2\Omega$ ($g^\prime \lesssim 10^{-8}$). As one can see from the plot, $g^\prime = 10^{-1}$ provides the strongest bound.

We conclude this section by noting that we do not perform a detailed analysis of the bounds on muonophilic light states. We only remark that very strong  bounds on light states are derived from fifth force searches. Most of these bounds do not apply in our case as these experiments are done using materials made out of protons, neutrons, and electrons.


\section{Conclusion}\label{sec:conc}

It is well known that fermion pairs can behave as bosons in several circumstances. In this work, we show that fermion pairs can also constitute classical radiation just like bosonic states do. We use this understanding to derive the generalization of the Larmor formula for the case of the fermion pair emission.

Being motivated by the potential of applying fermion pair radiation to astrophysical objects, we consider the case of classical sources following elliptical orbits. The most interesting regime of fermion pair radiation is when the mediator is off-shell, which takes place when the mass of the mediator is much smaller than the frequency of the periodic motion of the source. In this regime, the fermion pair emission takes over from on-shell boson production. This opens up a window into a broader region of parameter space for various models that allow for the fermion pair radiation by classical sources.

Subsequently, we apply our results to neutrino-antineutrino emission by two pulsar binary systems PSR B1913+16 and PSR J1738+0333. Neutrino pair emission by binary systems is highly suppressed in the SM compared to GW radiation, but can be significantly enhanced in various BSM scenarios. In particular, we consider two possibilities: light muonophilic vector and scalar mediators that couple to the SM neutrinos. Using period decay data for the two binary systems, we derive bounds on the parameters of the two models. While we did not perform a comprehensive study of the relevance of these bounds, the key point is that they provide a demonstration of the fact that fermion pair radiation can be used to enhance BSM probes using astrophysical data.

There are several future directions to go from here. Here are a few that we find particularly interesting:
\begin{itemize}
\item
A thorough and detailed study of the bounds that we find on specific models is called for. This, however, is complicated by the large uncertainties that come from the estimates on the neutron star constituents. In particular, new physics interactions alter the equation of state of a neutron star and, currently, there is no precise quantitative understanding of how this affects its content.
\item It also would be interesting to see if we can find more systems to which our results can be applied. In particular, exotic astrophysical systems and exotic types of new physics models.
\item
In this work, we only consider fermion pair radiation; however, the results can be modified to also include bosonic pair radiation. All that needs to be done is to calculate the relevant matrix elements. It is expected to result in a different kinematic dependence.
\end{itemize}

We conclude with the main message of our paper: If nature includes new light states, fermion pair radiation can be one more tool in our toolbox to probe them.

\section*{Acknowledgements} 
We are grateful to  
Kfir Blum,
Jeff Dror,
Toby Opferkuch, 
Nadav Outmezguine,
Ira Rothstein,
Ryosuke Sato, and
Kohsaku Tobioka
for useful discussions. 
The work of YG is supported in part by the NSF grant PHY1316222. The research of WT is supported by NSF Grant No. PHY-2013052.

\begin{appendix}\label{sec:app}

\section{Derivation of the power loss formula} 
\label{app:calc}


We present below an explicit derivation of the power loss formula for the fermion pair radiation by a point-like classical object on an elliptical orbit. We perform the calculation separately for the case of vector and scalar mediators. In our calculation, we follow closely the analysis in Ref.~\cite{KumarPoddar:2019ceq}.

\subsection{The case of a vector boson mediator}

The power loss is a sum over different harmonics, as given by Eqs.~\eqref{eq:dGamma2} and \eqref{eq:plosssummary}. The matrix element, at leading order, for a vector boson mediator, is given by Eq.~\eqref{eq:M-vector}. It includes the Fourier Transform of the classical current $J_{\text{cl}}^{\mu} (x)$ defined in Eq.~\eqref{eq:jcldef-ellipse}. We rewrite it here for convenience:

\beq
{\cal M}_n(s_1,s_2) =g^2 Q_{\psi}\,\bar{u}(k_1,s_1)\gamma^\mu v(k_2,s_2)\,\frac{i(-\eta_{\mu \nu}+(k_1+k_2)_{\mu}(k_1+k_2)_{\nu}/m_A^2)}{(k_1+k_2)^2-m_A^2 + i m_A \Gamma_A }\, J_{\text{cl}}^{\nu}(\Omega_n)\, ,
\eeq
where $\eta_{\mu\nu}$ is the Minkowski metric tensor.
Note that the contribution from the $(k_1 + k_2)_\mu (k_1 + k_2)_\nu$ term vanishes by means of the Dirac equation since the fermions are on-shell, that is,
\begin{eqnarray}
\bar{u}(\slashed{k}_{1}+\slashed{k}_{2})v=\bar{u}(m_\psi-m_\psi)v=0.
\end{eqnarray}

Squaring the amplitudes corresponding to different harmonics and summing over spins, we find
\begin{eqnarray}
	\label{eq:avgamp}
	|\overline{{\cal M}_n}|^2&=&\sum_{s_1,s_2}|{\cal M}_n|^2 = \frac{g^4 Q_\psi^2}{\left((k_1+k_2)^2-m_A^2\right)^2 + m_A^2 \Gamma_A^2}J_\text{cl}^\mu(\Omega_n) J_\text{cl}^{*\nu}(\Omega_n)\operatorname{Tr} \left[(\slashed{k}_{1}+m_{\nu})\gamma_{\mu}(\slashed{k}_{2}-m_\nu)\gamma_{\nu}\right] \nonumber\\
	&=&\frac{4g^4Q_\psi^2}{\left((k_1+k_2)^2-m_A^2\right)^2 + m_A^2 \Gamma_A^2}J_\text{cl}^\mu(\Omega_n) J_\text{cl}^{*\nu}(\Omega_n)\left(k_{1\mu} k_{2\nu} + k_{1\nu} k_{2\mu}-\frac{1}{2}(k_1+k_2)^2\eta_{\mu\nu}\right).
\end{eqnarray}

Finally, we are ready to write the expression for the rate of energy loss due to $\psi\bar{\psi}$ emission at harmonic $n$ by the classical source as
\begin{align}\label{eq:Pn}
	P_n = \left(\frac{dE}{dt}\right)_n &= \int \Omega_n \, d\Gamma_n \nonumber \\
	&= \Omega_n \int \frac{\D ^3{{\bm k}_1}}{(2\pi)^3(2\omega_1) }\frac{\D^3{\bm k}_2}{(2\pi)^3(2\omega_2) }(2\pi)\delta(\Omega_n-\omega_1-\omega_2)|\overline{\mathcal{M}_n}|^2\nonumber\\
	&= \Omega_n \int \D \Phi_1 \D\Phi_2\frac{|{\bm k}_1 | \D{\omega_1}}{2(2\pi)^3 }\frac{|{\bm k}_2 |\D{\omega_2}}{2(2\pi)^3 }(2\pi)\delta(\Omega_n-\omega_1-\omega_2)|\overline{\mathcal{M}_n}|^2 \, ,
\end{align}
where $|{\bm k}_{1,2}| =\sqrt{\omega_{1,2}^2 - m_\psi^2}$, we used $\Omega_n = \omega_1 + \omega_2$ for the total energy carried away by the fermion pair, $d \Phi_{1,2}$ are the differential elements of solid angles in the fermion's direction of flight, and $\left|\overline{\mathcal{M}_n}\right|^2$ is given in Eq.~\eqref{eq:avgamp}. The total power radiated is found by summing over all kinematically allowed harmonics:
\begin{equation}
    P = \sum_n P_n.
\end{equation}

To calculate the power radiated in fermion pairs by a point-like source in an elliptical orbit, we need to evaluate the integrals in Eq.~\eqref{eq:Pn}, after substituting in the explicit form of $J_\text{cl}^\mu(\Omega_n)$ in Eq.~\eqref{eq:avgamp}. Using Eqs.~\eqref{eq:jcldef-ellipse} and \eqref{eq:ellipse-param}, we find the Fourier Transform $J_{\text{cl}}^{\mu} (\Omega_n)$ as:
\begin{equation}\label{eq:J_Fourier}
    J_\text{cl}^i{(\Omega_n})= a\Omega Q j^i_n,\qquad J_\text{cl}^0(\Omega_n) =  a\Omega Q \left(\frac{{\bm j}_n\cdot {\bm p}}{n\Omega}\right),
\end{equation}
where the 3-vector ${\bm j}_n$ is defined as
\begin{equation}\label{eq:j-def}
 {\bm j}_n=\left(-iJ_n'(ne),\frac{\sqrt{1-e^2}}{e}J_n(ne),0\right),
\end{equation}
with $J_n(z)$ denoting a Bessel function, and ${\bm p} = {\bm k}_1 + {\bm k}_2 $.

The terms in the numerator of $|\overline{\mathcal{M}}|^2$ in Eq.~\eqref{eq:avgamp}, are then given by 
\begin{equation}\label{eq:JkJk}
    \left(J_\text{cl}^\mu (\Omega_n) k_{1\mu}\right)\left(J_\text{cl}^{\nu *} (\Omega_n) k_{2\nu}\right) = a^2  \Omega^2 Q^2 j_n^i j_n^{j*} \left[ \frac{\omega_1 \omega_2}{\left(n \Omega\right)^2}p^ip^j - \frac{\omega_1}{n\Omega}p^i k_2^j - \frac{\omega_2}{n\Omega}k_1^i p^j + k_1^i k_2^j\right],
\end{equation}
and
\begin{equation}\label{eq:J^2}
    |J^\mu_\text{cl}(\Omega_n)|^2 = |J_\text{cl}^0(\Omega_n)|^2 - |\bm{J}_\text{cl}(\Omega_n)|^2 = a^2  \Omega^2 Q^2 j_n^i j_n^{j*} \left[ \frac{p^i p^j}{(\Omega n)^2} - \delta^{ij}\right],
\end{equation}
where we used $\Omega_n = n\Omega$. Note that all quantities above are 3-vectors with Latin indices $i = 1,2,3$, and a sum over $i$ and $j$ is implicit. The expression for $\left(J_\text{cl}^\mu (\Omega_n) k_{2\mu}\right)\left(J_\text{cl}^{\nu*} (\Omega_n) k_{1\nu}\right)$ is obtained from Eq.~\eqref{eq:JkJk} via complex conjugation.

Next we note that the denominator of $|\overline{\mathcal{M}_n}|^2$, see Eq.~\eqref{eq:avgamp}, depends only on $m_A$, $\Gamma_A$, $\omega_{1,2}$, the magnitudes $|{\bm k}_{1,2}|$ and the relative angle between the two momenta ${\bm k}_{1}$, and ${\bm k}_{2}$ that we denote as $\gamma$. Because of this, it is convenient to perform the change of coordinates in the integral in Eq.~\eqref{eq:Pn} from the integration over the solid angles $\D \Phi_1 \D\Phi_2$ to the integration over $\D\Phi_1\D\Phi_2^{r}$ where the solid angle of the second neutrino is measured relative to the direction of ${\bm k}_1$, hence the super index $r$. (Equivalently, one can also choose to integrate over $\D\Phi_1^r \D\Phi_2$.) The Jacobian of this coordinate change is unity since the transformation is simply a coordinate rotation, and thus
\begin{equation}
    \D\Phi_1 \D\Phi_2 = \D\Phi_1 \D \Phi_2^r.
\end{equation}
Defining
\begin{equation}
    \D\Phi_b = \sin \theta_b \D \theta_b \D \phi_b, \qquad \D\Phi_2^r = \sin \gamma \D \gamma \D \delta, \qquad b = 1,2\,,
\end{equation}
we find the following relations between the two sets of integration variables
\begin{eqnarray}
	\cos \gamma&=&\cos \theta_1 \cos \theta_2+\sin \theta_1 \sin \theta_2 \cos \left(\phi_2-\phi_1\right), \nonumber\\
	\sin \delta&=&\frac{\sin \theta_2 \sin \left(\phi_2-\phi_1\right)}{\sin \gamma}\,.
\end{eqnarray}
Since, out of all the angular variables, the denominator only depends on the relative angle $\gamma$, the integrals over $\theta_1$, $\phi_1$ and $\delta$ can be taken easily using the following relations
\begin{eqnarray}\label{eq:ang-int-rel}
	\int \D\Phi_1 \D\Phi_2k_a^ik_a^j &= & \int \D \Phi_1 \D\Phi_2^r k_a^i k_a^j= \delta^{ij} \frac{8 \pi^2}{3} {\bm k}_a^2 \int \sin \gamma \D\gamma, 
	\nonumber \\
	\int \D\Phi_1 \D\Phi_2k_1^ik_2^j &=&  \int \D \Phi_1 \D\Phi_1^r k_1^ik_2^j = \delta^{ij} \frac{8\pi^2}{3} ({\bm k}_1 \cdot {\bm k}_2) \int \sin \gamma  \D \gamma, \nonumber \\
	\int \D\Phi_1 \D\Phi_2 &=& \int \D\Phi_1 \D\Phi_2^r = 8\pi^2 \int \sin\gamma \D\gamma\,.
\end{eqnarray}
Using this and the results of Eqs.~\eqref{eq:JkJk} and~\eqref{eq:J^2}, we perform the integration over $\theta_1$, $\phi_1$ and $\delta$ in Eq.~\eqref{eq:Pn}, and find the following expression for the power radiated in harmonic $n$,
\begin{align}\label{eq:Pn-gamma-w-w}
P_n &= \frac{g^4 \left(n \Omega\right)}{12 \pi^3}a^2 \Omega^2 Q_\psi^2 Q^2  \left|\bm{j}_n\right|^2
\int \frac{\delta(n\Omega -\omega_1 - \omega_2)}{\left((k_1+k_2)^2 -m_A^2\right)^2 + m_A^2 \Gamma_A^2} \times \nonumber \\
&\left[-\frac{1}{2}\left(k_1 + k_2 \right)^2 \left[\left({\bm k}_1 + {\bm k}_2\right)^2/\left(\left(n \Omega\right)^2 - 3 \right)\right] + 2 \frac{\omega_1 \omega_2}{\left( n \Omega \right)^2} \left({\bm k}_1 + {\bm k}_2\right)^2 
 \right. \nonumber\\ &
 \left. - 2 \frac{\omega_1}{n\Omega}\left({\bm k}_2^2 + {\bm k}_1 \cdot {\bm k}_2 \right) - 2 {\omega_2 \over n \Omega} \left({\bm k}_1^2+ {\bm k}_1 \cdot {\bm k}_2\right) + 2 {\bm k}_1 \cdot {\bm k}_2 \right]\times \nonumber\\ &
\omega_1 \omega_2 \left(1 - \frac{m_\psi^2}{\omega_1^2}\right)^{1/2}\left(1 - \frac{m_\psi^2}{\omega_2^2}\right)^{1/2} \sin \gamma \, \D\gamma \D \omega_1 \D \omega_2\,,
\end{align}
where the only integrals left are the integrals over $\gamma$, $\omega_1$ and $\omega_2$.

Next, we introduce the following dimensionless variables and parameters
\begin{equation} \label{eq:defns}
	x_1 = \frac{\omega_1}{\Omega}, \quad x_2 = \frac{\omega_2}{\Omega}, \quad n_\psi = \frac{m_\psi}{\Omega}, \quad n_A = \frac{m_A}{\Omega}, \quad n_\Gamma = \frac{\Gamma_A}{\Omega}.
\end{equation}
Performing the change of variables in Eq.~\eqref{eq:Pn-gamma-w-w} from $(\omega_1, \omega_2)$ to $(x_1, x_2)$, we rewrite the expression for the power radiated in harmonic $n$ as follows:
\begin{equation}
    P_n = \frac{g^4}{12 \pi^3}a^2  \Omega^4 Q_\psi ^2 |{\bm j}_n|^2
    \int \sin \gamma \, \D \gamma\ \D x_1 \, \D x_2\, \delta(n - x_1 - x_2)\, \mathcal{F}(\cos \gamma, x_1, x_2)\,.
\end{equation}
Upon taking the integral over $x_2$ and performing the replacement $x_1 \rightarrow x$, we obtain
\begin{equation}
    P_n = \frac{g^4}{12 \pi^3}a^2  \Omega^4 Q_\psi^2 Q^2 |{\bm j}_n|^2
    \int_{n_\psi}^{n-n_\psi} \D x \int_{-1}^1 \D (\cos \gamma)  \, \mathcal{F}(\cos \gamma, x)\,, \label{eq:def-calF-app}
\end{equation}
where function $\mathcal{F}(\cos \gamma, x)$ is given by 
\begin{equation}\label{eq:F-of-cos}
    \mathcal{F}(\cos \gamma, x) = \frac{b(x)}{2n}\frac{\frac{1}{2}b^2(x) \cos^2 \gamma + b(x)c(x) \cos \gamma + d(x)}{\left(a(x) - b(x) \cos \gamma\right)^2 + g^2}\,,
\end{equation}
with
\begin{eqnarray}
    a(x) &=& 2n_\psi^2 + 2x(n-x) -n_A^2\,,\nonumber \\
    b(x) &=& 2 \sqrt{x^2 - n_\psi^2}\sqrt{(n-x)^2 -n_\psi^2}\,, \nonumber\\
    c(x) & = & -\left(n^2 + 2n_\psi^2\right),\nonumber\\
    d(x) & = &  2 (x (n^3 - 2 n^2 x + 2 n x^2 - x^3) + 2 n^2 n_\psi^2 + n_\psi^4),\nonumber\\
    g^2 &=& n_A^2 n_\Gamma^2\,.
\end{eqnarray}
The variable $x$ here is the ratio of the energy of  one of the fermions to the fundamental oscillation frequency. It can be at least $n_\psi$ or at most $n - \npsi$, hence the limits on the integral. Also note that $F$ also depends on the parameters of the problem namely $n_A$, $n_\psi$, $n_\Gamma$ defined in Eq.~\eqref{eq:defns}, but we do not write them explicitly for brevity. Lastly, note that the $\gamma$-dependence of the numerator of function ${\cal F}$ is through a term quadratic in $\cos \gamma$ and a term linear in $\cos \gamma$. This behavior is attributed to the theory that we pick -- renormalizable theories such as in the case considered here would only contribute at most two powers of momentum in the matrix element, leading to a $\cos \gamma$ dependence that is at most quadratic. However non-renormalizable theories have more momenta in the matrix element, and will give us a different $\cos \gamma$ dependence in the ${\cal F}$.

Now, we define
\begin{equation}
    F^A(x) \equiv F ^A (n,x,\npsi, \na, \nga) = \int_{-1}^1 \D\left(\cos \gamma\right) \mathcal{F}\left(\cos \gamma, x,n\right), \label{eq:def-F-app}
\end{equation} 
where the superscript $A$ denotes the vector boson mediator.

The integral over $\cos \gamma$ can be taken analytically. Then, we find that the function $F^A(x)$, has the form:
\begin{eqnarray}
\label{eq:FM-gen}
F^A(x) = F^A_0 (x) &+&  {F^A_1(x)\over n_M \nga} \left[ \tan^{-1} \left({a(x)+b(x) \over n_M \nga }\right) - \tan^{-1} \left( {a(x)-b(x) \over n_M \nga }\right)\right] \nonumber \\
&+& F^A_2(x) \tanh ^{-1} \left ({2 a(x) b(x) \over  a(x)^2 + b(x)^2 + n_M^2 \nga^2}\right),
\end{eqnarray}
with:
\begin{eqnarray}
F^A_0(x) &=& b(x)/2n \, ,\nonumber \\
F^A_1(x)  &=& {1 \over 4n} \left(\na^4 + 4n^2 \npsi^2 - \na^2 \nga^2 + 2 \na^2 n^2 - 4nx\na^2 + 4x^2 \na^2 \right) \nonumber\, , \\
F^A_2(x) &=& {1 \over 2n} \left( \na^2 +n^2 -2 n x + 2x^2\right)\, .
\label{eq:f0f1f2b}
\end{eqnarray}
Consequently, the power loss formula of each mode with $n>2n_{\psi}$ becomes
\begin{eqnarray}
\label{power}
P_n=\frac{2 g^4 Q_\psi^2 Q^2 }{3(2\pi)^3}  a^2\Omega^4  \left(J^\prime_n(ne)^2 + \frac{1 - e^2}{e^2}J_n(ne)^2\right) \int_{n_{\psi}}^{n-n_{\psi}}\D x F^A(x),
\end{eqnarray}
which gives us Eq.~\eqref{eq:Ploss-A-mai} for the case $M =A$, where we define for mediator $M$
\beq \label{eq:Bn}
B_n^M(n_M,n_\nu,n_\Gamma)\,\equiv\,\left(J^\prime_n(ne)^2 + \frac{1 - e^2}{e^2}J_n(ne)^2\right) \int_{\npsi}^{n- \npsi} \D x \  F^{M}(x,n,n_M,n_\nu,n_\Gamma),
\eeq
where $J_n (z)$ is a Bessel function of order $n$ in the variable $z$.

\subsection{The case of the scalar mediator}\label{app:calc-scalar}
The derivation for the power loss in the scalar mediator is similar to the vector case, but the matrix element is different, as shown in Eq.~\eqref{eq:M-scalar}. This matrix element contains the number density $\rho_{\text{cl}} (x)$ of source particles, instead of a current. As such, the difference in the calculation in this case comes from the calculation of the squared matrix element, which in this case, is given by:
\begin{eqnarray}
	\sum_{s_1,s_2} |{\cal M}_n(s_1,s_2)|^2&=&\frac{g^2 g'^2 }{((k_1+k_2)^2-m_\phi^2)^2+m_\phi^2 \Gamma_\phi^2} \operatorname{Tr}((\slashed{k_1}+m_\psi)(\slashed{k_2}-m_\psi))|\rho_{\text{cl}}(\Omega_n)|^2\nonumber\\
	&=&\frac{4g^2 g'^2 }{((k_1+k_2)^2-m_\phi^2)^2+m_\phi^2 \Gamma_\phi^2}(k_1\cdot k_2-m_\psi^2)|\rho_{\text{cl}}(\Omega_n)|^2 \,].
	\end{eqnarray} 
The power loss is again given by Eq.~\eqref{eq:Pn}.

Using Eqs.~\eqref{eq:rhocldef-ellipse} and \eqref{eq:ellipse-param}, we find the Fourier Transform $\rho_{\text{cl}}^{\mu} (\Omega_n)$ as:
\begin{equation}\label{eq:rho_Fourier}
\rho_\text{cl}^0(\Omega_n) =  a\Omega N \left(\frac{{\bm j}_n\cdot {\bm p}}{n\Omega}\right),
\end{equation}
where, like in the vector case, we define the 3-vector $j_n^i$ as follows:
\begin{equation}\label{eq:j-def-phi}
 {\bm j}_n=\left(-iJ_n'(ne),\frac{\sqrt{1-e^2}}{e}J_n(ne),0\right),
\end{equation}
with $J_n(z)$ denoting a Bessel function, amd ${\bm p} = {\bm k}_1 + {\bm k}_2 $.

After performing all the steps analogous to Eqns.~\eqref{eq:Pn}--\eqref{eq:def-F-app} in the previous section, i.e, after performing the angular integration, we get:
\begin{equation}
    P_n = \frac{g^2 g'^2}{12 \pi^3}a^2  \Omega^4 N^2 |{\bm j}_n|^2
    \int_{n_\psi}^{n-n_\psi} \D x \int_{-1}^1 \D \cos \gamma  \, \mathcal{F}(\cos \gamma, x)\,,
\end{equation}
where function $\mathcal{F}(\cos \gamma, x)$ is given by 
\begin{equation}\label{eq:F-of-cos-phi}
    \mathcal{F}(\cos \gamma, x) = -\frac{b(x)}{2n}\frac{\frac{1}{2}b^2(x) \cos^2 \gamma + b(x)c(x) \cos \gamma + d(x)}{\left(a(x) - b(x) \cos \gamma\right)^2 + g^2}\,,
\end{equation}
with
\begin{eqnarray}
    a(x) &=& 2n_\psi^2 + 2x(n-x) -n_\phi^2\,,\nonumber \\
    b(x) &=& 2 \sqrt{x^2 - n_\psi^2}\sqrt{(n-x)^2 -n_\psi^2}\,, \nonumber\\
    c(x) & = & {(n-2x)^2 \over 2},\nonumber\\
    d(x) & = &  (\npsi^2 -nx +x^2)(n^2 - 2\npsi^2 -2nx+2x^2),\nonumber\\
    g^2 &=& n_\phi^2 n_\Gamma^2\,. 
\end{eqnarray}
Like before, we define
\begin{equation}
    F^\phi(x) \equiv F ^\phi (n,x,\npsi, \nphi, \nga) = \int_{-1}^1 \D \left(\cos \gamma\right) \mathcal{F}\left(\cos \gamma, x,n\right),
\end{equation} 
where the superscript $\phi$ denotes the scalar mediator.

The integral over $\cos \gamma$ can be taken analytically to find a form for $F^\phi$:
\begin{eqnarray}
F^\phi(x) = F^\phi_0 (x) &+&  {F^\phi_1(x)\over n_M \nga} \left[ \tan^{-1} \left({a(x)+b(x) \over n_M \nga }\right) - \tan^{-1} \left( {a(x)-b(x) \over n_M \nga }\right)\right] \nonumber \\
&+& F^\phi_2(x) \tanh ^{-1} \left ({2 a(x) b(x) \over  a(x)^2 + b(x)^2 + n_M^2 \nga^2}\right),
\end{eqnarray}
with:
\begin{eqnarray}\label{eq:f0f1f2scal}
F^\phi_0 (x) &=& -b(x)/2n \, ,\nonumber \\
F^\phi_1(x)  &=& {1 \over 4n} \left(\nphi^2 \nga^2 +(n^2-\nphi^2)(\nphi^2-4\nnu^2) \right) \, ,\nonumber \\
F^\phi_2(x) &=& {1 \over 4n} \left( n^2+4\nnu^2-2\nphi^2 \right) .
\end{eqnarray}
Consequently, the power loss formula of each mode with $n>2n_{\psi}$ becomes
\begin{eqnarray}
\label{power-phi}
P_n=\frac{2 g^2 g'^2 }{3(2\pi)^3}  a^2\Omega^4 N ^2 \left(J^\prime_n(ne)^2 + \frac{1 - e^2}{e^2}J_n(ne)^2\right) \int_{n_{\psi}}^{n-n_{\psi}} \D x F^\phi(x),
\end{eqnarray}
which gives us Eq.~\eqref{eq:Ploss-phi-mai} for the case $M =\phi$
\beq
P^{\phi}_n = \frac{g^2 g'^2 }{12 \pi^3}   a^2 \Omega^4 \left( {N_1 \over m_1} - {N_2 \over m_2}\right)^2 
B_n^\phi(n_A,n_\nu,n_\Gamma).\label{eq:Ploss-phi}
\eeq

We find that the form of the function $F^M$ is general for the two types of mediators, the difference lying in the explicit forms of the functions $F_0^M, F_1 ^M $ and $F_2^M$. This is due to the fact that the $\cos \gamma$ dependence of the function ${\cal F}$ is the same in both cases, as in both cases, the theory considered is a renormalizable one. As we explained in the previous sub-section, this general form of $F^M$ is not what we will have when we consider non-renormalizable theories that give us higher powers of momenta in the numerator of ${\cal F}$.
\end{appendix}

\bibliography{mybib}

\begin{thebibliography}{44}%
\makeatletter
\providecommand \@ifxundefined [1]{%
 \@ifx{#1\undefined}
}%
\providecommand \@ifnum [1]{%
 \ifnum #1\expandafter \@firstoftwo
 \else \expandafter \@secondoftwo
 \fi
}%
\providecommand \@ifx [1]{%
 \ifx #1\expandafter \@firstoftwo
 \else \expandafter \@secondoftwo
 \fi
}%
\providecommand \natexlab [1]{#1}%
\providecommand \enquote  [1]{``#1''}%
\providecommand \bibnamefont  [1]{#1}%
\providecommand \bibfnamefont [1]{#1}%
\providecommand \citenamefont [1]{#1}%
\providecommand \href@noop [0]{\@secondoftwo}%
\providecommand \href [0]{\begingroup \@sanitize@url \@href}%
\providecommand \@href[1]{\@@startlink{#1}\@@href}%
\providecommand \@@href[1]{\endgroup#1\@@endlink}%
\providecommand \@sanitize@url [0]{\catcode `\\12\catcode `\$12\catcode
  `\&12\catcode `\#12\catcode `\^12\catcode `\_12\catcode `\%12\relax}%
\providecommand \@@startlink[1]{}%
\providecommand \@@endlink[0]{}%
\providecommand \url  [0]{\begingroup\@sanitize@url \@url }%
\providecommand \@url [1]{\endgroup\@href {#1}{\urlprefix }}%
\providecommand \urlprefix  [0]{URL }%
\providecommand \Eprint [0]{\href }%
\providecommand \doibase [0]{http://dx.doi.org/}%
\providecommand \selectlanguage [0]{\@gobble}%
\providecommand \bibinfo  [0]{\@secondoftwo}%
\providecommand \bibfield  [0]{\@secondoftwo}%
\providecommand \translation [1]{[#1]}%
\providecommand \BibitemOpen [0]{}%
\providecommand \bibitemStop [0]{}%
\providecommand \bibitemNoStop [0]{.\EOS\space}%
\providecommand \EOS [0]{\spacefactor3000\relax}%
\providecommand \BibitemShut  [1]{\csname bibitem#1\endcsname}%
\let\auto@bib@innerbib\@empty
\bibitem [{\citenamefont {b.~Larmor}(1897)}]{LarmorLXIIIOT}%
  \BibitemOpen
  \bibfield  {author} {\bibinfo {author} {\bibfnamefont {J.~S.}\ \bibnamefont
  {b.~Larmor}},\ }\href@noop {} {\bibfield  {journal} {\bibinfo  {journal}
  {Philosophical Magazine Series 1}\ }\textbf {\bibinfo {volume} {44}},\
  \bibinfo {pages} {503} (\bibinfo {year} {1897})}\BibitemShut {NoStop}%
\bibitem [{\citenamefont {Jackson}(1998)}]{Jackson:1998nia}%
  \BibitemOpen
  \bibfield  {author} {\bibinfo {author} {\bibfnamefont {J.~D.}\ \bibnamefont
  {Jackson}},\ }\href@noop {} {\emph {\bibinfo {title} {{Classical
  Electrodynamics}}}}\ (\bibinfo  {publisher} {Wiley},\ \bibinfo {year}
  {1998})\BibitemShut {NoStop}%
\bibitem [{\citenamefont {Krause}\ \emph {et~al.}(1994)\citenamefont {Krause},
  \citenamefont {Kloor},\ and\ \citenamefont {Fischbach}}]{Krause:1994ar}%
  \BibitemOpen
  \bibfield  {author} {\bibinfo {author} {\bibfnamefont {D.}~\bibnamefont
  {Krause}}, \bibinfo {author} {\bibfnamefont {H.~T.}\ \bibnamefont {Kloor}}, \
  and\ \bibinfo {author} {\bibfnamefont {E.}~\bibnamefont {Fischbach}},\ }\href
  {\doibase 10.1103/PhysRevD.49.6892} {\bibfield  {journal} {\bibinfo
  {journal} {Phys. Rev. D}\ }\textbf {\bibinfo {volume} {49}},\ \bibinfo
  {pages} {6892} (\bibinfo {year} {1994})}\BibitemShut {NoStop}%
\bibitem [{\citenamefont {Mohanty}\ and\ \citenamefont
  {Kumar~Panda}(1996)}]{Mohanty:1994yi}%
  \BibitemOpen
  \bibfield  {author} {\bibinfo {author} {\bibfnamefont {S.}~\bibnamefont
  {Mohanty}}\ and\ \bibinfo {author} {\bibfnamefont {P.}~\bibnamefont
  {Kumar~Panda}},\ }\href {\doibase 10.1103/PhysRevD.53.5723} {\bibfield
  {journal} {\bibinfo  {journal} {Phys. Rev. D}\ }\textbf {\bibinfo {volume}
  {53}},\ \bibinfo {pages} {5723} (\bibinfo {year} {1996})},\ \Eprint
  {http://arxiv.org/abs/hep-ph/9403205} {arXiv:hep-ph/9403205} \BibitemShut
  {NoStop}%
\bibitem [{\citenamefont {Dror}\ \emph {et~al.}(2020)\citenamefont {Dror},
  \citenamefont {Laha},\ and\ \citenamefont {Opferkuch}}]{Dror:2019uea}%
  \BibitemOpen
  \bibfield  {author} {\bibinfo {author} {\bibfnamefont {J.~A.}\ \bibnamefont
  {Dror}}, \bibinfo {author} {\bibfnamefont {R.}~\bibnamefont {Laha}}, \ and\
  \bibinfo {author} {\bibfnamefont {T.}~\bibnamefont {Opferkuch}},\ }\href
  {\doibase 10.1103/PhysRevD.102.023005} {\bibfield  {journal} {\bibinfo
  {journal} {Phys. Rev. D}\ }\textbf {\bibinfo {volume} {102}},\ \bibinfo
  {pages} {023005} (\bibinfo {year} {2020})},\ \Eprint
  {http://arxiv.org/abs/1909.12845} {arXiv:1909.12845 [hep-ph]} \BibitemShut
  {NoStop}%
\bibitem [{\citenamefont {Kumar~Poddar}\ \emph {et~al.}(2019)\citenamefont
  {Kumar~Poddar}, \citenamefont {Mohanty},\ and\ \citenamefont
  {Jana}}]{KumarPoddar:2019ceq}%
  \BibitemOpen
  \bibfield  {author} {\bibinfo {author} {\bibfnamefont {T.}~\bibnamefont
  {Kumar~Poddar}}, \bibinfo {author} {\bibfnamefont {S.}~\bibnamefont
  {Mohanty}}, \ and\ \bibinfo {author} {\bibfnamefont {S.}~\bibnamefont
  {Jana}},\ }\href {\doibase 10.1103/PhysRevD.100.123023} {\bibfield  {journal}
  {\bibinfo  {journal} {Phys. Rev. D}\ }\textbf {\bibinfo {volume} {100}},\
  \bibinfo {pages} {123023} (\bibinfo {year} {2019})},\ \Eprint
  {http://arxiv.org/abs/1908.09732} {arXiv:1908.09732 [hep-ph]} \BibitemShut
  {NoStop}%
\bibitem [{\citenamefont {Huang}\ \emph {et~al.}(2019)\citenamefont {Huang},
  \citenamefont {Johnson}, \citenamefont {Sagunski}, \citenamefont
  {Sakellariadou},\ and\ \citenamefont {Zhang}}]{Huang:2018pbu}%
  \BibitemOpen
  \bibfield  {author} {\bibinfo {author} {\bibfnamefont {J.}~\bibnamefont
  {Huang}}, \bibinfo {author} {\bibfnamefont {M.~C.}\ \bibnamefont {Johnson}},
  \bibinfo {author} {\bibfnamefont {L.}~\bibnamefont {Sagunski}}, \bibinfo
  {author} {\bibfnamefont {M.}~\bibnamefont {Sakellariadou}}, \ and\ \bibinfo
  {author} {\bibfnamefont {J.}~\bibnamefont {Zhang}},\ }\href {\doibase
  10.1103/PhysRevD.99.063013} {\bibfield  {journal} {\bibinfo  {journal} {Phys.
  Rev. D}\ }\textbf {\bibinfo {volume} {99}},\ \bibinfo {pages} {063013}
  (\bibinfo {year} {2019})},\ \Eprint {http://arxiv.org/abs/1807.02133}
  {arXiv:1807.02133 [hep-ph]} \BibitemShut {NoStop}%
\bibitem [{\citenamefont {Kumar~Poddar}\ \emph {et~al.}(2020)\citenamefont
  {Kumar~Poddar}, \citenamefont {Mohanty},\ and\ \citenamefont
  {Jana}}]{KumarPoddar:2019jxe}%
  \BibitemOpen
  \bibfield  {author} {\bibinfo {author} {\bibfnamefont {T.}~\bibnamefont
  {Kumar~Poddar}}, \bibinfo {author} {\bibfnamefont {S.}~\bibnamefont
  {Mohanty}}, \ and\ \bibinfo {author} {\bibfnamefont {S.}~\bibnamefont
  {Jana}},\ }\href {\doibase 10.1103/PhysRevD.101.083007} {\bibfield  {journal}
  {\bibinfo  {journal} {Phys. Rev. D}\ }\textbf {\bibinfo {volume} {101}},\
  \bibinfo {pages} {083007} (\bibinfo {year} {2020})},\ \Eprint
  {http://arxiv.org/abs/1906.00666} {arXiv:1906.00666 [hep-ph]} \BibitemShut
  {NoStop}%
\bibitem [{\citenamefont {Hook}\ and\ \citenamefont
  {Huang}(2018)}]{Hook:2017psm}%
  \BibitemOpen
  \bibfield  {author} {\bibinfo {author} {\bibfnamefont {A.}~\bibnamefont
  {Hook}}\ and\ \bibinfo {author} {\bibfnamefont {J.}~\bibnamefont {Huang}},\
  }\href {\doibase 10.1007/JHEP06(2018)036} {\bibfield  {journal} {\bibinfo
  {journal} {JHEP}\ }\textbf {\bibinfo {volume} {06}},\ \bibinfo {pages} {036}
  (\bibinfo {year} {2018})},\ \Eprint {http://arxiv.org/abs/1708.08464}
  {arXiv:1708.08464 [hep-ph]} \BibitemShut {NoStop}%
\bibitem [{\citenamefont {Foot}(1991)}]{Foot:1990mn}%
  \BibitemOpen
  \bibfield  {author} {\bibinfo {author} {\bibfnamefont {R.}~\bibnamefont
  {Foot}},\ }\href {\doibase 10.1142/S0217732391000543} {\bibfield  {journal}
  {\bibinfo  {journal} {Mod. Phys. Lett. A}\ }\textbf {\bibinfo {volume} {6}},\
  \bibinfo {pages} {527} (\bibinfo {year} {1991})}\BibitemShut {NoStop}%
\bibitem [{\citenamefont {He}\ \emph {et~al.}(1991)\citenamefont {He},
  \citenamefont {Joshi}, \citenamefont {Lew},\ and\ \citenamefont
  {Volkas}}]{He:1991qd}%
  \BibitemOpen
  \bibfield  {author} {\bibinfo {author} {\bibfnamefont {X.-G.}\ \bibnamefont
  {He}}, \bibinfo {author} {\bibfnamefont {G.~C.}\ \bibnamefont {Joshi}},
  \bibinfo {author} {\bibfnamefont {H.}~\bibnamefont {Lew}}, \ and\ \bibinfo
  {author} {\bibfnamefont {R.~R.}\ \bibnamefont {Volkas}},\ }\href {\doibase
  10.1103/PhysRevD.44.2118} {\bibfield  {journal} {\bibinfo  {journal} {Phys.
  Rev. D}\ }\textbf {\bibinfo {volume} {44}},\ \bibinfo {pages} {2118}
  (\bibinfo {year} {1991})}\BibitemShut {NoStop}%
\bibitem [{\citenamefont {Foot}\ \emph {et~al.}(1994)\citenamefont {Foot},
  \citenamefont {He}, \citenamefont {Lew},\ and\ \citenamefont
  {Volkas}}]{Foot:1994vd}%
  \BibitemOpen
  \bibfield  {author} {\bibinfo {author} {\bibfnamefont {R.}~\bibnamefont
  {Foot}}, \bibinfo {author} {\bibfnamefont {X.~G.}\ \bibnamefont {He}},
  \bibinfo {author} {\bibfnamefont {H.}~\bibnamefont {Lew}}, \ and\ \bibinfo
  {author} {\bibfnamefont {R.~R.}\ \bibnamefont {Volkas}},\ }\href {\doibase
  10.1103/PhysRevD.50.4571} {\bibfield  {journal} {\bibinfo  {journal} {Phys.
  Rev. D}\ }\textbf {\bibinfo {volume} {50}},\ \bibinfo {pages} {4571}
  (\bibinfo {year} {1994})},\ \Eprint {http://arxiv.org/abs/hep-ph/9401250}
  {arXiv:hep-ph/9401250} \BibitemShut {NoStop}%
\bibitem [{\citenamefont {Heeck}\ and\ \citenamefont
  {Rodejohann}(2011)}]{Heeck:2011wj}%
  \BibitemOpen
  \bibfield  {author} {\bibinfo {author} {\bibfnamefont {J.}~\bibnamefont
  {Heeck}}\ and\ \bibinfo {author} {\bibfnamefont {W.}~\bibnamefont
  {Rodejohann}},\ }\href {\doibase 10.1103/PhysRevD.84.075007} {\bibfield
  {journal} {\bibinfo  {journal} {Phys. Rev. D}\ }\textbf {\bibinfo {volume}
  {84}},\ \bibinfo {pages} {075007} (\bibinfo {year} {2011})},\ \Eprint
  {http://arxiv.org/abs/1107.5238} {arXiv:1107.5238 [hep-ph]} \BibitemShut
  {NoStop}%
\bibitem [{\citenamefont {Kopp}\ \emph {et~al.}(2018)\citenamefont {Kopp},
  \citenamefont {Laha}, \citenamefont {Opferkuch},\ and\ \citenamefont
  {Shepherd}}]{Kopp:2018jom}%
  \BibitemOpen
  \bibfield  {author} {\bibinfo {author} {\bibfnamefont {J.}~\bibnamefont
  {Kopp}}, \bibinfo {author} {\bibfnamefont {R.}~\bibnamefont {Laha}}, \bibinfo
  {author} {\bibfnamefont {T.}~\bibnamefont {Opferkuch}}, \ and\ \bibinfo
  {author} {\bibfnamefont {W.}~\bibnamefont {Shepherd}},\ }\href {\doibase
  10.1007/JHEP11(2018)096} {\bibfield  {journal} {\bibinfo  {journal} {JHEP}\
  }\textbf {\bibinfo {volume} {11}},\ \bibinfo {pages} {096} (\bibinfo {year}
  {2018})},\ \Eprint {http://arxiv.org/abs/1807.02527} {arXiv:1807.02527
  [hep-ph]} \BibitemShut {NoStop}%
\bibitem [{\citenamefont {Alexander}\ \emph {et~al.}(2018)\citenamefont
  {Alexander}, \citenamefont {McDonough}, \citenamefont {Sims},\ and\
  \citenamefont {Yunes}}]{Alexander:2018qzg}%
  \BibitemOpen
  \bibfield  {author} {\bibinfo {author} {\bibfnamefont {S.}~\bibnamefont
  {Alexander}}, \bibinfo {author} {\bibfnamefont {E.}~\bibnamefont
  {McDonough}}, \bibinfo {author} {\bibfnamefont {R.}~\bibnamefont {Sims}}, \
  and\ \bibinfo {author} {\bibfnamefont {N.}~\bibnamefont {Yunes}},\ }\href
  {\doibase 10.1088/1361-6382/aaeb5c} {\bibfield  {journal} {\bibinfo
  {journal} {Class. Quant. Grav.}\ }\textbf {\bibinfo {volume} {35}},\ \bibinfo
  {pages} {235012} (\bibinfo {year} {2018})},\ \Eprint
  {http://arxiv.org/abs/1808.05286} {arXiv:1808.05286 [gr-qc]} \BibitemShut
  {NoStop}%
\bibitem [{\citenamefont {Choi}\ and\ \citenamefont
  {Jung}(2019)}]{Choi:2018axi}%
  \BibitemOpen
  \bibfield  {author} {\bibinfo {author} {\bibfnamefont {H.~G.}\ \bibnamefont
  {Choi}}\ and\ \bibinfo {author} {\bibfnamefont {S.}~\bibnamefont {Jung}},\
  }\href {\doibase 10.1103/PhysRevD.99.015013} {\bibfield  {journal} {\bibinfo
  {journal} {Phys. Rev. D}\ }\textbf {\bibinfo {volume} {99}},\ \bibinfo
  {pages} {015013} (\bibinfo {year} {2019})},\ \Eprint
  {http://arxiv.org/abs/1810.01421} {arXiv:1810.01421 [hep-ph]} \BibitemShut
  {NoStop}%
\bibitem [{\citenamefont {Fabbrichesi}\ and\ \citenamefont
  {Urbano}(2020)}]{Fabbrichesi:2019ema}%
  \BibitemOpen
  \bibfield  {author} {\bibinfo {author} {\bibfnamefont {M.}~\bibnamefont
  {Fabbrichesi}}\ and\ \bibinfo {author} {\bibfnamefont {A.}~\bibnamefont
  {Urbano}},\ }\href {\doibase 10.1088/1475-7516/2020/06/007} {\bibfield
  {journal} {\bibinfo  {journal} {JCAP}\ }\textbf {\bibinfo {volume} {06}},\
  \bibinfo {pages} {007} (\bibinfo {year} {2020})},\ \Eprint
  {http://arxiv.org/abs/1902.07914} {arXiv:1902.07914 [hep-ph]} \BibitemShut
  {NoStop}%
\bibitem [{\citenamefont {Seymour}\ and\ \citenamefont
  {Yagi}(2020)}]{Seymour:2019tir}%
  \BibitemOpen
  \bibfield  {author} {\bibinfo {author} {\bibfnamefont {B.~C.}\ \bibnamefont
  {Seymour}}\ and\ \bibinfo {author} {\bibfnamefont {K.}~\bibnamefont {Yagi}},\
  }\href {\doibase 10.1088/1361-6382/ab9933} {\bibfield  {journal} {\bibinfo
  {journal} {Class. Quant. Grav.}\ }\textbf {\bibinfo {volume} {37}},\ \bibinfo
  {pages} {145008} (\bibinfo {year} {2020})},\ \Eprint
  {http://arxiv.org/abs/1908.03353} {arXiv:1908.03353 [gr-qc]} \BibitemShut
  {NoStop}%
\bibitem [{\citenamefont {Feinberg}\ and\ \citenamefont
  {Sucher}(1968)}]{Feinberg:1968zz}%
  \BibitemOpen
  \bibfield  {author} {\bibinfo {author} {\bibfnamefont {G.}~\bibnamefont
  {Feinberg}}\ and\ \bibinfo {author} {\bibfnamefont {J.}~\bibnamefont
  {Sucher}},\ }\href {\doibase 10.1103/PhysRev.166.1638} {\bibfield  {journal}
  {\bibinfo  {journal} {Phys. Rev.}\ }\textbf {\bibinfo {volume} {166}},\
  \bibinfo {pages} {1638} (\bibinfo {year} {1968})}\BibitemShut {NoStop}%
\bibitem [{\citenamefont {Hsu}\ and\ \citenamefont
  {Sikivie}(1994)}]{Hsu:1992tg}%
  \BibitemOpen
  \bibfield  {author} {\bibinfo {author} {\bibfnamefont {S.~D.~H.}\
  \bibnamefont {Hsu}}\ and\ \bibinfo {author} {\bibfnamefont {P.}~\bibnamefont
  {Sikivie}},\ }\href {\doibase 10.1103/PhysRevD.49.4951} {\bibfield  {journal}
  {\bibinfo  {journal} {Phys. Rev. D}\ }\textbf {\bibinfo {volume} {49}},\
  \bibinfo {pages} {4951} (\bibinfo {year} {1994})},\ \Eprint
  {http://arxiv.org/abs/hep-ph/9211301} {arXiv:hep-ph/9211301} \BibitemShut
  {NoStop}%
\bibitem [{\citenamefont {Ghosh}\ \emph {et~al.}(2020)\citenamefont {Ghosh},
  \citenamefont {Grossman},\ and\ \citenamefont {Tangarife}}]{Ghosh:2019dmi}%
  \BibitemOpen
  \bibfield  {author} {\bibinfo {author} {\bibfnamefont {M.}~\bibnamefont
  {Ghosh}}, \bibinfo {author} {\bibfnamefont {Y.}~\bibnamefont {Grossman}}, \
  and\ \bibinfo {author} {\bibfnamefont {W.}~\bibnamefont {Tangarife}},\ }\href
  {\doibase 10.1103/PhysRevD.101.116006} {\bibfield  {journal} {\bibinfo
  {journal} {Phys. Rev. D}\ }\textbf {\bibinfo {volume} {101}},\ \bibinfo
  {pages} {116006} (\bibinfo {year} {2020})},\ \Eprint
  {http://arxiv.org/abs/1912.09444} {arXiv:1912.09444 [hep-ph]} \BibitemShut
  {NoStop}%
\bibitem [{\citenamefont {Ghosh}\ \emph {et~al.}(2022)\citenamefont {Ghosh},
  \citenamefont {Grossman}, \citenamefont {Tangarife}, \citenamefont {Xu},\
  and\ \citenamefont {Yu}}]{Ghosh:2022nzo}%
  \BibitemOpen
  \bibfield  {author} {\bibinfo {author} {\bibfnamefont {M.}~\bibnamefont
  {Ghosh}}, \bibinfo {author} {\bibfnamefont {Y.}~\bibnamefont {Grossman}},
  \bibinfo {author} {\bibfnamefont {W.}~\bibnamefont {Tangarife}}, \bibinfo
  {author} {\bibfnamefont {X.-J.}\ \bibnamefont {Xu}}, \ and\ \bibinfo {author}
  {\bibfnamefont {B.}~\bibnamefont {Yu}},\ }\href@noop {} {\  (\bibinfo {year}
  {2022})},\ \Eprint {http://arxiv.org/abs/2209.07082} {arXiv:2209.07082
  [hep-ph]} \BibitemShut {NoStop}%
\bibitem [{\citenamefont {van Straten}\ \emph {et~al.}(2001)\citenamefont {van
  Straten}, \citenamefont {Bailes}, \citenamefont {Britton}, \citenamefont
  {Kulkarni}, \citenamefont {Anderson}, \citenamefont {Manchester},\ and\
  \citenamefont {Sarkissian}}]{vanStraten:2001zk}%
  \BibitemOpen
  \bibfield  {author} {\bibinfo {author} {\bibfnamefont {W.}~\bibnamefont {van
  Straten}}, \bibinfo {author} {\bibfnamefont {M.}~\bibnamefont {Bailes}},
  \bibinfo {author} {\bibfnamefont {M.~C.}\ \bibnamefont {Britton}}, \bibinfo
  {author} {\bibfnamefont {S.~R.}\ \bibnamefont {Kulkarni}}, \bibinfo {author}
  {\bibfnamefont {S.~B.}\ \bibnamefont {Anderson}}, \bibinfo {author}
  {\bibfnamefont {R.~N.}\ \bibnamefont {Manchester}}, \ and\ \bibinfo {author}
  {\bibfnamefont {J.}~\bibnamefont {Sarkissian}},\ }\href {\doibase
  10.1038/35084015} {\bibfield  {journal} {\bibinfo  {journal} {Nature}\
  }\textbf {\bibinfo {volume} {412}},\ \bibinfo {pages} {158} (\bibinfo {year}
  {2001})},\ \Eprint {http://arxiv.org/abs/astro-ph/0108254}
  {arXiv:astro-ph/0108254} \BibitemShut {NoStop}%
\bibitem [{\citenamefont {Kramer}\ \emph {et~al.}(2006)\citenamefont {Kramer}
  \emph {et~al.}}]{Kramer:2006nb}%
  \BibitemOpen
  \bibfield  {author} {\bibinfo {author} {\bibfnamefont {M.}~\bibnamefont
  {Kramer}} \emph {et~al.},\ }\href {\doibase 10.1126/science.1132305}
  {\bibfield  {journal} {\bibinfo  {journal} {Science}\ }\textbf {\bibinfo
  {volume} {314}},\ \bibinfo {pages} {97} (\bibinfo {year} {2006})},\ \Eprint
  {http://arxiv.org/abs/astro-ph/0609417} {arXiv:astro-ph/0609417} \BibitemShut
  {NoStop}%
\bibitem [{\citenamefont {Stairs}\ \emph {et~al.}(2002)\citenamefont {Stairs},
  \citenamefont {Thorsett}, \citenamefont {Taylor},\ and\ \citenamefont
  {Wolszczan}}]{Stairs:2002cw}%
  \BibitemOpen
  \bibfield  {author} {\bibinfo {author} {\bibfnamefont {I.~H.}\ \bibnamefont
  {Stairs}}, \bibinfo {author} {\bibfnamefont {S.~E.}\ \bibnamefont
  {Thorsett}}, \bibinfo {author} {\bibfnamefont {J.~H.}\ \bibnamefont
  {Taylor}}, \ and\ \bibinfo {author} {\bibfnamefont {A.}~\bibnamefont
  {Wolszczan}},\ }\href {\doibase 10.1086/344157} {\bibfield  {journal}
  {\bibinfo  {journal} {Astrophys. J.}\ }\textbf {\bibinfo {volume} {581}},\
  \bibinfo {pages} {501} (\bibinfo {year} {2002})},\ \Eprint
  {http://arxiv.org/abs/astro-ph/0208357} {arXiv:astro-ph/0208357} \BibitemShut
  {NoStop}%
\bibitem [{\citenamefont {Shannon}\ \emph {et~al.}(2014)\citenamefont
  {Shannon}, \citenamefont {Johnston},\ and\ \citenamefont
  {Manchester}}]{Shannon:2013dpa}%
  \BibitemOpen
  \bibfield  {author} {\bibinfo {author} {\bibfnamefont {R.~M.}\ \bibnamefont
  {Shannon}}, \bibinfo {author} {\bibfnamefont {S.}~\bibnamefont {Johnston}}, \
  and\ \bibinfo {author} {\bibfnamefont {R.~N.}\ \bibnamefont {Manchester}},\
  }\href {\doibase 10.1093/mnras/stt2123} {\bibfield  {journal} {\bibinfo
  {journal} {Mon. Not. Roy. Astron. Soc.}\ }\textbf {\bibinfo {volume} {437}},\
  \bibinfo {pages} {3255} (\bibinfo {year} {2014})},\ \Eprint
  {http://arxiv.org/abs/1311.0588} {arXiv:1311.0588 [astro-ph.SR]} \BibitemShut
  {NoStop}%
\bibitem [{\citenamefont {Antoniadis}\ \emph {et~al.}(2013)\citenamefont
  {Antoniadis} \emph {et~al.}}]{Antoniadis:2013pzd}%
  \BibitemOpen
  \bibfield  {author} {\bibinfo {author} {\bibfnamefont {J.}~\bibnamefont
  {Antoniadis}} \emph {et~al.},\ }\href {\doibase 10.1126/science.1233232}
  {\bibfield  {journal} {\bibinfo  {journal} {Science}\ }\textbf {\bibinfo
  {volume} {340}},\ \bibinfo {pages} {6131} (\bibinfo {year} {2013})},\ \Eprint
  {http://arxiv.org/abs/1304.6875} {arXiv:1304.6875 [astro-ph.HE]} \BibitemShut
  {NoStop}%
\bibitem [{\citenamefont {Bhat}\ \emph {et~al.}(2008)\citenamefont {Bhat},
  \citenamefont {Bailes},\ and\ \citenamefont {Verbiest}}]{Bhat:2008ck}%
  \BibitemOpen
  \bibfield  {author} {\bibinfo {author} {\bibfnamefont {N.~D.~R.}\
  \bibnamefont {Bhat}}, \bibinfo {author} {\bibfnamefont {M.}~\bibnamefont
  {Bailes}}, \ and\ \bibinfo {author} {\bibfnamefont {J.~P.~W.}\ \bibnamefont
  {Verbiest}},\ }\href {\doibase 10.1103/PhysRevD.77.124017} {\bibfield
  {journal} {\bibinfo  {journal} {Phys. Rev. D}\ }\textbf {\bibinfo {volume}
  {77}},\ \bibinfo {pages} {124017} (\bibinfo {year} {2008})},\ \Eprint
  {http://arxiv.org/abs/0804.0956} {arXiv:0804.0956 [astro-ph]} \BibitemShut
  {NoStop}%
\bibitem [{\citenamefont {Freire}\ \emph {et~al.}(2012)\citenamefont {Freire},
  \citenamefont {Wex}, \citenamefont {Esposito-Farese}, \citenamefont
  {Verbiest}, \citenamefont {Bailes}, \citenamefont {Jacoby}, \citenamefont
  {Kramer}, \citenamefont {Stairs}, \citenamefont {Antoniadis},\ and\
  \citenamefont {Janssen}}]{Freire:2012mg}%
  \BibitemOpen
  \bibfield  {author} {\bibinfo {author} {\bibfnamefont {P.~C.~C.}\
  \bibnamefont {Freire}}, \bibinfo {author} {\bibfnamefont {N.}~\bibnamefont
  {Wex}}, \bibinfo {author} {\bibfnamefont {G.}~\bibnamefont
  {Esposito-Farese}}, \bibinfo {author} {\bibfnamefont {J.~P.~W.}\ \bibnamefont
  {Verbiest}}, \bibinfo {author} {\bibfnamefont {M.}~\bibnamefont {Bailes}},
  \bibinfo {author} {\bibfnamefont {B.~A.}\ \bibnamefont {Jacoby}}, \bibinfo
  {author} {\bibfnamefont {M.}~\bibnamefont {Kramer}}, \bibinfo {author}
  {\bibfnamefont {I.~H.}\ \bibnamefont {Stairs}}, \bibinfo {author}
  {\bibfnamefont {J.}~\bibnamefont {Antoniadis}}, \ and\ \bibinfo {author}
  {\bibfnamefont {G.~H.}\ \bibnamefont {Janssen}},\ }\href {\doibase
  10.1111/j.1365-2966.2012.21253.x} {\bibfield  {journal} {\bibinfo  {journal}
  {Mon. Not. Roy. Astron. Soc.}\ }\textbf {\bibinfo {volume} {423}},\ \bibinfo
  {pages} {3328} (\bibinfo {year} {2012})},\ \Eprint
  {http://arxiv.org/abs/1205.1450} {arXiv:1205.1450 [astro-ph.GA]} \BibitemShut
  {NoStop}%
\bibitem [{\citenamefont {Ferdman}\ \emph {et~al.}(2014)\citenamefont {Ferdman}
  \emph {et~al.}}]{Ferdman:2014rna}%
  \BibitemOpen
  \bibfield  {author} {\bibinfo {author} {\bibfnamefont {R.~D.}\ \bibnamefont
  {Ferdman}} \emph {et~al.},\ }\href {\doibase 10.1093/mnras/stu1223}
  {\bibfield  {journal} {\bibinfo  {journal} {Mon. Not. Roy. Astron. Soc.}\
  }\textbf {\bibinfo {volume} {443}},\ \bibinfo {pages} {2183} (\bibinfo {year}
  {2014})},\ \Eprint {http://arxiv.org/abs/1406.5507} {arXiv:1406.5507
  [astro-ph.SR]} \BibitemShut {NoStop}%
\bibitem [{\citenamefont {van Leeuwen}\ \emph {et~al.}(2015)\citenamefont {van
  Leeuwen} \emph {et~al.}}]{vanLeeuwen:2014sca}%
  \BibitemOpen
  \bibfield  {author} {\bibinfo {author} {\bibfnamefont {J.}~\bibnamefont {van
  Leeuwen}} \emph {et~al.},\ }\href {\doibase 10.1088/0004-637X/798/2/118}
  {\bibfield  {journal} {\bibinfo  {journal} {Astrophys. J.}\ }\textbf
  {\bibinfo {volume} {798}},\ \bibinfo {pages} {118} (\bibinfo {year}
  {2015})},\ \Eprint {http://arxiv.org/abs/1411.1518} {arXiv:1411.1518
  [astro-ph.SR]} \BibitemShut {NoStop}%
\bibitem [{\citenamefont {Jacoby}\ \emph {et~al.}(2006)\citenamefont {Jacoby},
  \citenamefont {Cameron}, \citenamefont {Jenet}, \citenamefont {Anderson},
  \citenamefont {Murty},\ and\ \citenamefont {Kulkarni}}]{Jacoby:2006dy}%
  \BibitemOpen
  \bibfield  {author} {\bibinfo {author} {\bibfnamefont {B.~A.}\ \bibnamefont
  {Jacoby}}, \bibinfo {author} {\bibfnamefont {P.~B.}\ \bibnamefont {Cameron}},
  \bibinfo {author} {\bibfnamefont {F.~A.}\ \bibnamefont {Jenet}}, \bibinfo
  {author} {\bibfnamefont {S.~B.}\ \bibnamefont {Anderson}}, \bibinfo {author}
  {\bibfnamefont {R.~N.}\ \bibnamefont {Murty}}, \ and\ \bibinfo {author}
  {\bibfnamefont {S.~R.}\ \bibnamefont {Kulkarni}},\ }\href {\doibase
  10.1086/505742} {\bibfield  {journal} {\bibinfo  {journal} {Astrophys. J.
  Lett.}\ }\textbf {\bibinfo {volume} {644}},\ \bibinfo {pages} {L113}
  (\bibinfo {year} {2006})},\ \Eprint {http://arxiv.org/abs/astro-ph/0605375}
  {arXiv:astro-ph/0605375} \BibitemShut {NoStop}%
\bibitem [{\citenamefont {Davoudiasl}\ and\ \citenamefont
  {Denton}(2019)}]{Davoudiasl:2019nlo}%
  \BibitemOpen
  \bibfield  {author} {\bibinfo {author} {\bibfnamefont {H.}~\bibnamefont
  {Davoudiasl}}\ and\ \bibinfo {author} {\bibfnamefont {P.~B.}\ \bibnamefont
  {Denton}},\ }\href {\doibase 10.1103/PhysRevLett.123.021102} {\bibfield
  {journal} {\bibinfo  {journal} {Phys. Rev. Lett.}\ }\textbf {\bibinfo
  {volume} {123}},\ \bibinfo {pages} {021102} (\bibinfo {year} {2019})},\
  \Eprint {http://arxiv.org/abs/1904.09242} {arXiv:1904.09242 [astro-ph.CO]}
  \BibitemShut {NoStop}%
\bibitem [{\citenamefont {Hulse}\ and\ \citenamefont
  {Taylor}(1975)}]{Hulse:1974eb}%
  \BibitemOpen
  \bibfield  {author} {\bibinfo {author} {\bibfnamefont {R.~A.}\ \bibnamefont
  {Hulse}}\ and\ \bibinfo {author} {\bibfnamefont {J.~H.}\ \bibnamefont
  {Taylor}},\ }\href {\doibase 10.1086/181708} {\bibfield  {journal} {\bibinfo
  {journal} {Astrophys. J. Lett.}\ }\textbf {\bibinfo {volume} {195}},\
  \bibinfo {pages} {L51} (\bibinfo {year} {1975})}\BibitemShut {NoStop}%
\bibitem [{\citenamefont {Taylor}\ and\ \citenamefont
  {Weisberg}(1982)}]{Taylor:1982zz}%
  \BibitemOpen
  \bibfield  {author} {\bibinfo {author} {\bibfnamefont {J.~H.}\ \bibnamefont
  {Taylor}}\ and\ \bibinfo {author} {\bibfnamefont {J.~M.}\ \bibnamefont
  {Weisberg}},\ }\href {\doibase 10.1086/159690} {\bibfield  {journal}
  {\bibinfo  {journal} {Astrophys. J.}\ }\textbf {\bibinfo {volume} {253}},\
  \bibinfo {pages} {908} (\bibinfo {year} {1982})}\BibitemShut {NoStop}%
\bibitem [{\citenamefont {Weisberg}\ and\ \citenamefont
  {Huang}(2016)}]{weisberg2016relativistic}%
  \BibitemOpen
  \bibfield  {author} {\bibinfo {author} {\bibfnamefont {J.~M.}\ \bibnamefont
  {Weisberg}}\ and\ \bibinfo {author} {\bibfnamefont {Y.}~\bibnamefont
  {Huang}},\ }\href {\doibase 10.3847/0004-637X/829/1/55} {\bibfield  {journal}
  {\bibinfo  {journal} {The Astrophysical Journal}\ }\textbf {\bibinfo {volume}
  {829}},\ \bibinfo {pages} {55} (\bibinfo {year} {2016})}\BibitemShut
  {NoStop}%
\bibitem [{\citenamefont {Kilic}\ \emph {et~al.}(2015)\citenamefont {Kilic},
  \citenamefont {Hermes}, \citenamefont {Gianninas},\ and\ \citenamefont
  {Brown}}]{kilic2015psr}%
  \BibitemOpen
  \bibfield  {author} {\bibinfo {author} {\bibfnamefont {M.}~\bibnamefont
  {Kilic}}, \bibinfo {author} {\bibfnamefont {J.}~\bibnamefont {Hermes}},
  \bibinfo {author} {\bibfnamefont {A.}~\bibnamefont {Gianninas}}, \ and\
  \bibinfo {author} {\bibfnamefont {W.~R.}\ \bibnamefont {Brown}},\ }\href
  {\doibase 10.1093/mnrasl/slu152} {\bibfield  {journal} {\bibinfo  {journal}
  {Monthly Notices of the Royal Astronomical Society: Letters}\ }\textbf
  {\bibinfo {volume} {446}},\ \bibinfo {pages} {L26} (\bibinfo {year}
  {2015})}\BibitemShut {NoStop}%
\bibitem [{\citenamefont {Peters}\ and\ \citenamefont
  {Mathews}(1963)}]{Peters:1963ux}%
  \BibitemOpen
  \bibfield  {author} {\bibinfo {author} {\bibfnamefont {P.~C.}\ \bibnamefont
  {Peters}}\ and\ \bibinfo {author} {\bibfnamefont {J.}~\bibnamefont
  {Mathews}},\ }\href {\doibase 10.1103/PhysRev.131.435} {\bibfield  {journal}
  {\bibinfo  {journal} {Phys. Rev.}\ }\textbf {\bibinfo {volume} {131}},\
  \bibinfo {pages} {435} (\bibinfo {year} {1963})}\BibitemShut {NoStop}%
\bibitem [{\citenamefont {Yakovlev}\ \emph {et~al.}(2001)\citenamefont
  {Yakovlev}, \citenamefont {Kaminker}, \citenamefont {Gnedin},\ and\
  \citenamefont {Haensel}}]{Yakovlev:2000jp}%
  \BibitemOpen
  \bibfield  {author} {\bibinfo {author} {\bibfnamefont {D.~G.}\ \bibnamefont
  {Yakovlev}}, \bibinfo {author} {\bibfnamefont {A.~D.}\ \bibnamefont
  {Kaminker}}, \bibinfo {author} {\bibfnamefont {O.~Y.}\ \bibnamefont
  {Gnedin}}, \ and\ \bibinfo {author} {\bibfnamefont {P.}~\bibnamefont
  {Haensel}},\ }\href {\doibase 10.1016/S0370-1573(00)00131-9} {\bibfield
  {journal} {\bibinfo  {journal} {Phys. Rept.}\ }\textbf {\bibinfo {volume}
  {354}},\ \bibinfo {pages} {1} (\bibinfo {year} {2001})},\ \Eprint
  {http://arxiv.org/abs/astro-ph/0012122} {arXiv:astro-ph/0012122} \BibitemShut
  {NoStop}%
\bibitem [{\citenamefont {Garani}\ and\ \citenamefont
  {Heeck}(2019)}]{Garani:2019fpa}%
  \BibitemOpen
  \bibfield  {author} {\bibinfo {author} {\bibfnamefont {R.}~\bibnamefont
  {Garani}}\ and\ \bibinfo {author} {\bibfnamefont {J.}~\bibnamefont {Heeck}},\
  }\href {\doibase 10.1103/PhysRevD.100.035039} {\bibfield  {journal} {\bibinfo
   {journal} {Phys. Rev. D}\ }\textbf {\bibinfo {volume} {100}},\ \bibinfo
  {pages} {035039} (\bibinfo {year} {2019})},\ \Eprint
  {http://arxiv.org/abs/1906.10145} {arXiv:1906.10145 [hep-ph]} \BibitemShut
  {NoStop}%
\bibitem [{\citenamefont {Pearson}\ \emph {et~al.}(2018)\citenamefont
  {Pearson}, \citenamefont {Chamel}, \citenamefont {Potekhin}, \citenamefont
  {Fantina}, \citenamefont {Ducoin}, \citenamefont {Dutta},\ and\ \citenamefont
  {Goriely}}]{Pearson:2018tkr}%
  \BibitemOpen
  \bibfield  {author} {\bibinfo {author} {\bibfnamefont {J.~M.}\ \bibnamefont
  {Pearson}}, \bibinfo {author} {\bibfnamefont {N.}~\bibnamefont {Chamel}},
  \bibinfo {author} {\bibfnamefont {A.~Y.}\ \bibnamefont {Potekhin}}, \bibinfo
  {author} {\bibfnamefont {A.~F.}\ \bibnamefont {Fantina}}, \bibinfo {author}
  {\bibfnamefont {C.}~\bibnamefont {Ducoin}}, \bibinfo {author} {\bibfnamefont
  {A.~K.}\ \bibnamefont {Dutta}}, \ and\ \bibinfo {author} {\bibfnamefont
  {S.}~\bibnamefont {Goriely}},\ }\href {\doibase 10.1093/mnras/sty2413}
  {\bibfield  {journal} {\bibinfo  {journal} {Mon. Not. Roy. Astron. Soc.}\
  }\textbf {\bibinfo {volume} {481}},\ \bibinfo {pages} {2994} (\bibinfo {year}
  {2018})},\ \bibinfo {note} {[Erratum: Mon.Not.Roy.Astron.Soc. 486, 768
  (2019)]},\ \Eprint {http://arxiv.org/abs/1903.04981} {arXiv:1903.04981
  [astro-ph.HE]} \BibitemShut {NoStop}%
\bibitem [{\citenamefont {Harry}\ and\ \citenamefont
  {Hinderer}(2018)}]{Harry:2018hke}%
  \BibitemOpen
  \bibfield  {author} {\bibinfo {author} {\bibfnamefont {I.}~\bibnamefont
  {Harry}}\ and\ \bibinfo {author} {\bibfnamefont {T.}~\bibnamefont
  {Hinderer}},\ }\href {\doibase 10.1088/1361-6382/aac7e3} {\bibfield
  {journal} {\bibinfo  {journal} {Class. Quant. Grav.}\ }\textbf {\bibinfo
  {volume} {35}},\ \bibinfo {pages} {145010} (\bibinfo {year} {2018})},\
  \Eprint {http://arxiv.org/abs/1801.09972} {arXiv:1801.09972 [gr-qc]}
  \BibitemShut {NoStop}%
\bibitem [{\citenamefont {Zhang}\ and\ \citenamefont
  {Chen}(2001)}]{Zhang:2000my}%
  \BibitemOpen
  \bibfield  {author} {\bibinfo {author} {\bibfnamefont {F.}~\bibnamefont
  {Zhang}}\ and\ \bibinfo {author} {\bibfnamefont {L.}~\bibnamefont {Chen}},\
  }\href {\doibase 10.1088/0256-307X/18/1/350} {\bibfield  {journal} {\bibinfo
  {journal} {Chin. Phys. Lett.}\ }\textbf {\bibinfo {volume} {18}},\ \bibinfo
  {pages} {142} (\bibinfo {year} {2001})},\ \Eprint
  {http://arxiv.org/abs/nucl-th/0011017} {arXiv:nucl-th/0011017} \BibitemShut
  {NoStop}%
\bibitem [{\citenamefont {Maki}\ \emph {et~al.}(1962)\citenamefont {Maki},
  \citenamefont {Nakagawa},\ and\ \citenamefont {Sakata}}]{Maki:1962mu}%
  \BibitemOpen
  \bibfield  {author} {\bibinfo {author} {\bibfnamefont {Z.}~\bibnamefont
  {Maki}}, \bibinfo {author} {\bibfnamefont {M.}~\bibnamefont {Nakagawa}}, \
  and\ \bibinfo {author} {\bibfnamefont {S.}~\bibnamefont {Sakata}},\ }\href
  {\doibase 10.1143/PTP.28.870} {\bibfield  {journal} {\bibinfo  {journal}
  {Prog. Theor. Phys.}\ }\textbf {\bibinfo {volume} {28}},\ \bibinfo {pages}
  {870} (\bibinfo {year} {1962})}\BibitemShut {NoStop}%
\end{thebibliography}%
\bibliographystyle{apsrev4-1}

\end{document}